%% file: paper.tex
\documentclass[a4paper]{article}
\usepackage{graphicx}
\usepackage{twocolceurws}
\usepackage{cite}
\usepackage{amsmath,amssymb,amsfonts}
\usepackage{algorithm}
\usepackage{textcomp}
\usepackage[caption = false]{subfig}
\usepackage{hyperref}
\usepackage{hhline}
\usepackage{makecell}
\usepackage{algpseudocode}
\usepackage{multirow}
\usepackage{tabularx}
\usepackage{stfloats}

\usepackage{bm}
\usepackage[font=scriptsize,labelfont=bf]{caption}
\usepackage{color}
\definecolor{lightgray}{gray}{0.9}

\newcommand\greybox[1]{%
  \medskip
  \par\noindent\colorbox{lightgray}{%
    \begin{minipage}{0.47\textwidth}\ttfamily\footnotesize #1\end{minipage}%
  }%
  \medskip%
}

\title{Creative Agents: Simulating the Systems Model of Creativity with Generative Agents}

\author{
Naomi Imasato$^{[1]}$ 
        \\ \scriptsize c.imasato@rlg.sys.es.osaka-u.ac.jp
\and
Kazuki Miyazawa$^{[1]}$
\and
Takayuki Nagai$^{[1,2]}$
\and
Takato Horii$^{[1,3]}$
}

\institution{$^{[1]}$Graduate School of Engineering Science, Osaka University, Osaka, Japan \\
$^{[2]}$Artificial Intelligence Exploration Research Center, The University of Electro-Communications, Tokyo, Japan \\
$^{[3]}$International Research Center for Neurointelligence (WPI-IRCN), The University of Tokyo, Tokyo, Japan}

\begin{document}
\maketitle

\begin{abstract}
With the growing popularity of generative AI for images, video, and music, we witnessed models rapidly improve
in quality and performance. However, not much attention is paid towards enabling AI's ability to ``be creative''.
In this study, we implemented and simulated the systems
model of creativity (proposed by Csikszentmihalyi) using
virtual agents utilizing large language models (LLMs) and text prompts. For comparison, the simulations were conducted with the  ``virtual artists'' being: 1)isolated and 
2)placed in a multi-agent system. Both scenarios were compared by analyzing the variations
and overall ``creativity'' in the generated artifacts 
(measured via a user study and LLM). Our results suggest that the generative agents may
perform better in the framework of the systems model of creativity.
\end{abstract}

\section{Introduction}
\input{tex/1Introduction}


\section{Related works}
\input{tex/2Related_works.tex}
  

\section{Method}
\label{method}
\input{tex/3Method}


\section{Experiments}
\input{tex/4Experiments}


\section{Results}
\input{tex/5Results}

\section{Discussion}
\input{tex/6Discussion}


\section{Conclusion}
\input{tex/7Conclusion}

\section{Appendix A\break Progression of text prompts}
\label{appendix:prompts}
\input{tex/AAppendix}

\section{Appendix B\break Impact of the text prompt}
\label{appendix:impact}
\input{tex/BAppendix}

\bibliographystyle{alpha} 
\bibliography{manual_bib,references}

\end{document}

%% file: tex/1Introduction.tex
\begin{figure}
    \centering
    \includegraphics[scale=0.34]{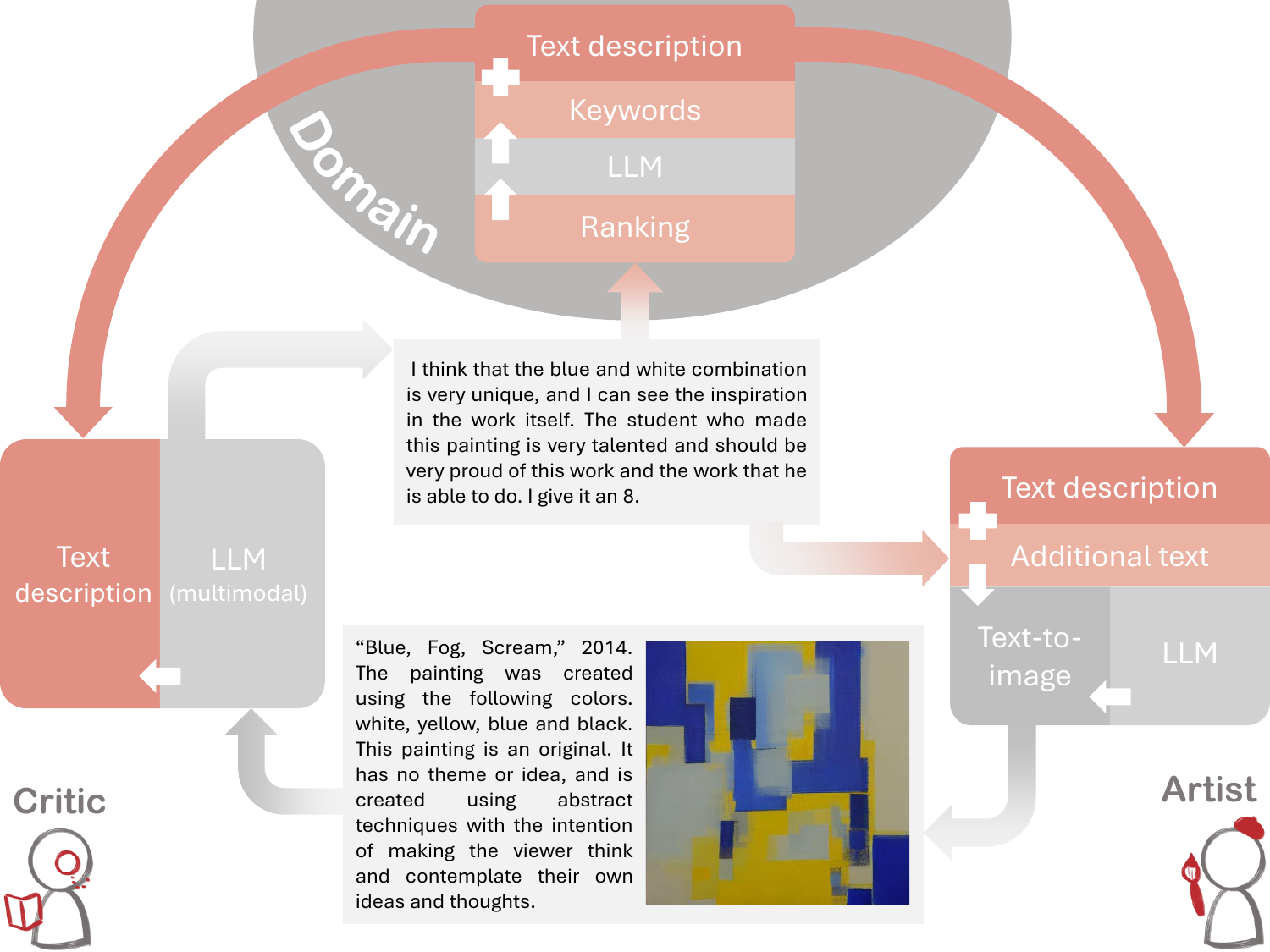}
    \caption{Each part of the system is composed by a combination of pieces of text and generative models. 
    The \textit{Domain} is described in text form, which is passed to complement the descriptions for both \textit{Artist}
    and \textit{Critic}. The text description in the \textit{Artist} is used in a text prompt template that generated the
    text that describes the \textit{Artist}'s new piece of art (art prompt). The art prompt is used to generate the image
    that should depict what the \textit{Artist} described. Both the image and the art prompt are fed into an LLM, where
    another template for a text prompt is filled with the text description for the \textit{Critic}. This will give us the
    critique that this \textit{Critic} wants to give to the \textit{Artist}'s latest artwork. This critique is used to
    place this new artwork in the overall ranking of the system, which represents changes in the \textit{Domain}. The same
    critique is also fed into the LLM with the \textit{Artist}'s text description in another text prompt template to
    make the \textit{Artist} ``reflect'' over the critique it just received. The output from this ``reflection'' goes
    into the \textit{Artist}'s additional text. We run these steps iteratively, where the text defining the \textit{Artist}
    and the ranking in the \textit{Domain} change at each step.}
    \label{fig:overview}
\end{figure}
  
  As the use of artificial intelligence (AI) models and solutions is becoming common,
researchers and AI enthusiasts have been working on maximally utilizing
the definition of \textit{``Intelligence''} in AI. There is an increasing interest in
exploring a machine capability for solving problems that are not well-defined, which requires
being creative.
  
  The definition of creativity varies across different sources of literature, but two factors that are consistently
referred to are \textit{novelty} and \textit{value}, both of which are difficult to define in a vacuum,
as \textit{novelty} refers to something being presented for the first time, and \textit{value} refers to the importance
or relevance that it holds \cite{amabile1996creativity,runco2012standard}. The complexity in defining
\textit{novelty} and \textit{value} lies in the fact that both imply a collective meaning. To determine
\textit{novelty}, the relative creations must be known, as well as to whom they were presented to. In practice,
the public relies on the information and commentaries propagated by specialists. For example, if an art curator
indicates that a certain painting is credibly unique and has a high market value, we are
compelled to believe it. Artists seeking to create new artwork may also rely on the implications
of this aforementioned curator to remain updated with the current standards of creativity,
thus have a higher likelihood of producing a successful artwork. This dynamic was described in
the work of Csikszentmihalyi. He proposed a systems model of creativity \cite{Csikszentmihalyi2015TheCsikszentmihalyi}
in which the society is partitioned into three main groups. Each part plays a role in determining and shaping the standards
for what is considered ``novel'' and ``valuable''; namely, ``creative''.

%% file: tex/2Related_works.tex
\subsection{Creativity}
  A fundamental problem in the field of computational
creativity is defining \textit{creativity}. Researchers have used different approaches
to define creativity. 
Multiple studies by Boden are often used as references to define creativity
\cite{Boden2004TheMechanisms,Boden2009ComputerCreativity}.
Boden often defined a hypothetical space of ideas, the process of forming
new ideas in this space, and whether these new ideas indicate a transformation in the space of concern.
Despite its popularity, Boden's definition poses a major obstacle for computational implementation,
as it requires high levels of abstraction to represent all the ``ideas'' in a space.
Conversely, Csikszentmihalyi designed a systems model of creativity, 
which proposes that creativity is product of social interactions and 
transforms overtime as part of societal change \cite{Csikszentmihalyi2015TheCsikszentmihalyi}.
Unlike Boden's definition, Csikszentmihalyi's approach considers the process of making something
``creative'' as a product of the interactions between multiple parts, which can be more
easily achieved with the current technology.

  We based our study on the definition of Csikszentmihalyi's
systems model of creativity; therefore, in this study, we will not focus on determining the creativity
in a single individual, rather observing the impact of social interactions on ``creativity'' in a
virtual environment with multiple generative agents.

\begin{figure}[H]
    \centering
    \includegraphics[width=\linewidth]{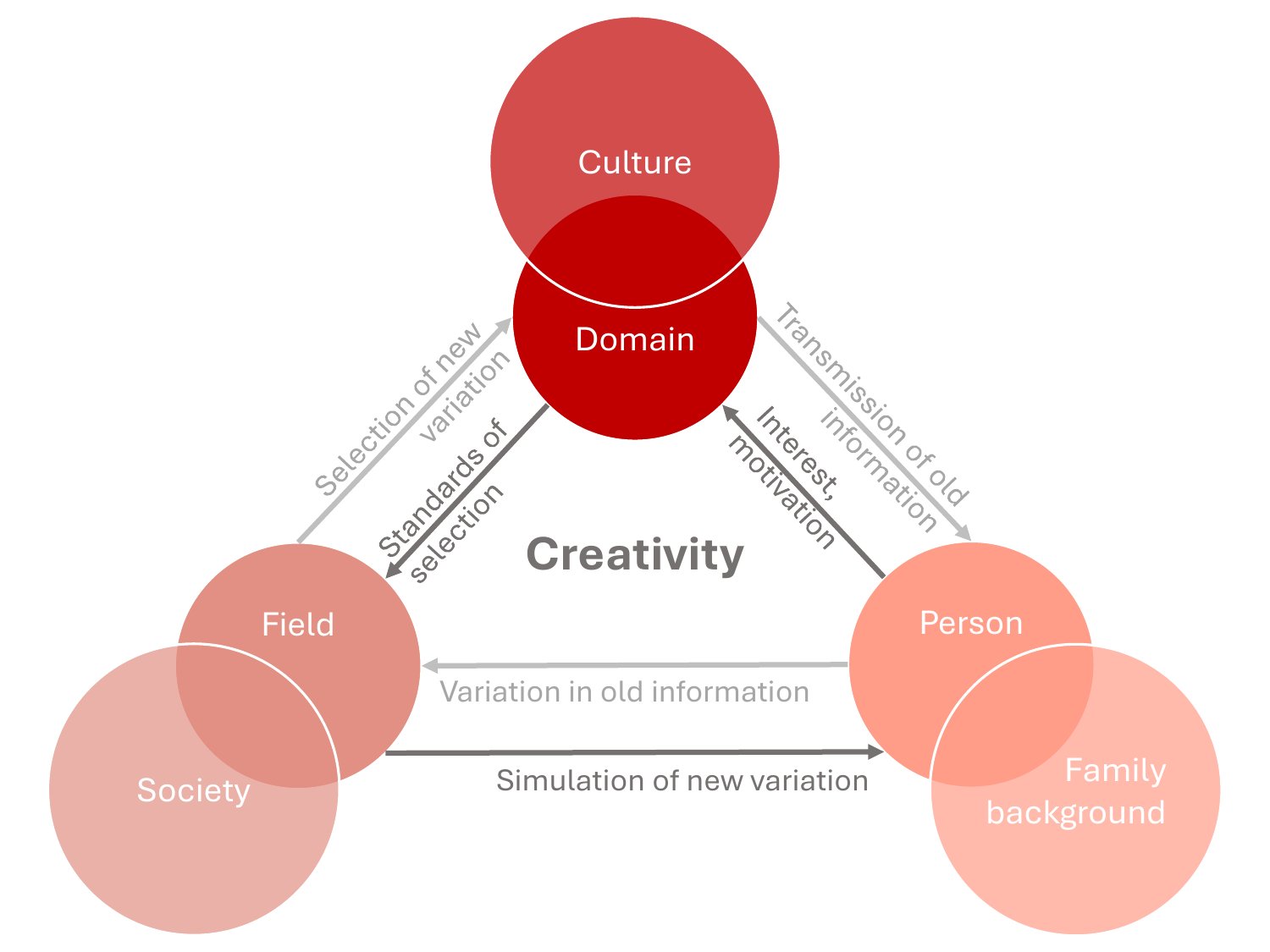}
    \caption{Diagram of the systems model of creativity as proposed by Csikszentmihalyi.}
    \label{fig:systemsmodel}
\end{figure}

  The systems model of creativity features individuals who create artifacts (i.e.
artists), the community to whom the artifact is being displayed, and the context in which the artifact
was created (e.g. location, historical events, significant pieces of art, recent trends and 
political climate). Each part of the system that does not live in a vacuum is constantly affecting
the others and vice-versa: the artist does not create art without inspiration or motivation
from the community or context they are in, the community cannot exist without multiple individuals
contributing to the domain with new artifacts, and the domain cannot change without the creations
of the artists or the judgment of the community. All the variables form a cycle in which each part
of the system contributes to the system itself. A simple diagram of 
the systems model of creativity is shown in Figure \ref{fig:systemsmodel}.

\subsection{Creativity and AI}
  Researchers in the field of computational creativity have attempted to develop
a computer program capable of being creative, such as the Joke Analysis
and Production Engine (Jape) \cite{Binsted1996MachinePuns}. It was developed at a time where AI was at
its early stages of development, and was designed to create puns. Humor heavily relies on creativity,
considering its surprising aspect that is found by many.
The author used \textit{exploratory programming}, which indicates that the
program was used to explore and test ideas. They designed a formal model for punning riddles based
on patterns observed in the structures and mechanisms of the indicated riddles; this formal model was
then used to implement Jape. The riddles that were used as a base for the analysis were obtained from children's
books that contained a collection of jokes. Despite the author claiming that there were ``a huge number 
of riddles to choose from'', which may have been the case at that time, three books
is relatively limited for the current standards of a dataset. To implement the model, they stored the information
regarding the phonological, semantic, syntactic and surface forms (appearance, spelling) of the words, as well as the
relationship between these lexical items. These were then applied to templates, where the relationships were
properly placed in the form of punning riddles. Although this is a robust model for generating punning riddles,
which is better than merely picking words to fill a riddle template, the use of these templates
restricts and limits the possibilities of outputs from this model. Moreover, significantly more effort is required to prepare
the program for creating new riddles compared to that when using the modern generative models.

  Generative adversarial networks (GANs) \cite{Goodfellow2014GenerativeNetworks} were once a popular choice
for computational creativity owing to the latent space created during training (alluding to Boden's
hypothetical ``space of ideas''). 
GANs are formed by using a generator and discriminator model, where both parts are 
trained to outdo the other (hence, ``adversarial'');
DesIGN is an example of this model \cite{sbai2018design}. The authors proposed StyleGAN, an adapted GAN
model where the generator is conditioned by the texture and the shape of the articles of clothing while
being ``encouraged'' to generate more novel samples, as opposed to replicating the samples observed in training,
as that performed by the classic GAN. This ``encouragement'' was achieved by adding two loss
functions that ``confuse'' the discriminator in the GAN. However, GANs are notoriously difficult to
train, as both the discriminator and generator should ideally perform sufficiently; however, achieving a balanced
training process is challenging.

  Transformer-based models have recently become a staple of generative models 
\cite{Vaswani2017AttentionNeed,Huang2018MusicTransformer,Radford2019LanguageLearners}. 
Considering the development of transformers, researchers have proposed a method for 
controlled text generation where the user can specify 
``tags'' or ``keywords'' to manipulate the probability distribution of the next word to be sampled
by an autoregressive model (GPT-2). For example, the Plug and Play Language Model (PPLM) \cite{Dathathri2019PlugGeneration},
successfully controlled the generation
of text sequences, where a significant portion of the outputs followed the ``tag'' or ``keyword''
specified without compromising the output coherence or grammatical accuracy. Another noteworthy
experiment conducted by the authors utilized more than one ``tag'' or ``keyword'' for the
controlled text generation where the control terms were unlikely pairs. 
Although the authors did not assess the creative ability of this method, they
shared some noteworthy results obtained by combining different themes and concepts.
Compared to Jape, the model imposed significantly milder constraints and limits,
and the outputs were based on a large volume of written information sourced
from the Web. The limit is based on the ability of the model to tie two or more concepts together when
they are significantly different from one another.
We successfully used this controlled generation method
in music generation \cite{Imasato2023UsingEmotion}, where
the autoregressive model was trained to generate sequences of MIDI events instead of text. Control was achieved using a part of the
PPLM method, where we used an auxiliary discriminator model to classify the data into two or more categories.
In this case, the emotion perceived from a piece of music was used as the main factor (categorized according to Russell's model of
affect). Analogous to the original publication of PPLM, using unlikely pairs of controls and prompts can lead to noteworthy
(or creative) outputs.

  Assessing or measuring the creativity of an artifact is one of the hardest (if not the hardest) problems in this field.
As humans, we visualize or hear a certain type of art or music, respectively, and almost immediately
make judgments based on the first impressions and our preferences. If desired, we also seek for
context, debate and further discussions with other people regarding whether the art or music of concern can be considered creative.
However, if asked to define measurements or objective values to determine the creativity of an
artifact, the reply is significantly less trivial and more complex than determining whether 
something or someone is creative. In an experiment, the authors of Jape considered 122 children who read jokes 
(generated by Jape and humans) and non-jokes (sensible and nonsensible ones). The
texts generated by Jape performed almost as well as the human-made
jokes compared to the non-joke texts \cite{Binsted1996MachinePuns}. 
To assess the creativity and funniness of a punning riddle, the authors
resorted to the decisions made by a group of individuals who participated in their experiment; as the judgment of
whether something is creative or funny is highly subjective, this is a significantly common practice among the different
works in the field. Certain objective measurements can be considered, such as the choice of words for poetry, color
palettes for images, among others. However these metrics do not determine creativity; in fact, many artifacts
can have good values for these metrics and remain to be considered non-creative, similar to one with poor
values in the same metrics being considered highly creative. In this regard, 
the quality of computationally generated artifacts can be easily measured with objective metrics; however measuring their 
creativity is significantly more complex given the subjectivity of the matter.

\subsection{Human-likeness in AI models}
  Modelling human behavior to better understand
the mechanisms and dynamics in our society is a major goal of AI research. In this regard, large language models (LLMs) integrated in chat bots,
such as GPT-4 integrated in ChatGPT \cite{openai2024gpt4}, are among the most notable developments
in the field. The quality of outputs obtained by ChatGPT users is significantly high, leading to 
discussions regarding Artificial General Intelligence (AGI) and ethics of the training and usage
of these AI models and systems.

  A recent study used LLMs to simulate individual agents in a small community, where each agent was
defined by text excerpts used to prompt text generation for their actions and utterances
\cite{Park2023GenerativeBehavior}. The authors also conducted a study
to assess the human-likeness of the agents, as well as the generation of actions and phrases by LLMs that
appeared natural to most participants. Additionally, when an agent was initialized with a
plan to host an event for the community, the agent invited other agents to the
event, who did the same and invited other agents. 

  Another study used a similar principle but maintained the LLM requests to a minimum, thus required a lower demand of
resources \cite{Kaiya2023LyfeInteractions}. The authors also used different ``situations'' where the agents
would have specific problems to solve to evaluate their performance. These studies indicate 
that LLM-based agents can fulfill their roles in a human-like manner and successfully
coordinate among themselves as a community.

  As LLMs were successfully used in simulating ``individuals'' in both of the aforementioned studies, we hypothesized that
LLM-based agents can be used to run simulations of the Csikszentmihalyi's systems model of
creativity \cite{Csikszentmihalyi2015TheCsikszentmihalyi}.

%% file: tex/3Method.tex
  In this study, we designed, implemented and tested a simplified simulation of the systems model
of creativity, as proposed by Csikszentmihalyi \cite{Csikszentmihalyi2015TheCsikszentmihalyi}.
In this section, we describe the overall structure of our system and
the decisions made during its implementation.

\subsection{The Systems Model of Creativity}
  According to Csikszentmihalyi \cite{Csikszentmihalyi2015TheCsikszentmihalyi},
creativity does not originate from an object or person in isolation; rather, 
\textit{``it is the product of three main shaping forces: a set
of social institutions, or field, that selects from the variations produced by individuals those
that are worth preserving; a stable cultural domain that will preserve and transmit the selected new
ideas or forms to the following generations; and finally the individual, who brings about some change
in the domain, a change that the field, will consider to be creative. (…) Creativity is a phenomenon
that results from the interaction between these three systems.''}.
Based on this description, three essential elements must be considered in a systems model
of creativity: the creative individual (or simply \textit{individual}), field and domain. Note,
at the end of the aforementioned passage, the author refers to each part as a \textit{``system''}, as 
each part is highly complex on its own.

  The creative individual would not be the same if it was not for the context they are in. This includes
(but is not limited to) the political, historical, and cultural contexts they are surround by.
Although people have their own individuality, personality, and experiences,
their surroundings play a major role in shaping their opinions and values. 
Despite using text, describing each detail of an individual and 
``modeling'' them would be practically impossible. A single individual
is difficult to model precisely. Modeling the field and domain is an even
more complicated task, as they are formed and affected by multiple agents, each with their
own complexities.

\begin{algorithm}
\caption{Simulation of systems model of creativity}
\label{alg:sys}
\begin{algorithmic}[1]
\State Initialize \textit{artists, field, domain}
\State $t=0$
\While{$t < 15$}
\For{\textit{artist} in \textit{artists}} 
    \State prompt $\gets$ art\_templ(\textit{domain}.desc,\textit{artist}.desc)
    \State \textit{artist}.art\_prompt $\gets$ LLM(prompt)
    \State \textit{artist}.image $\gets$ text2img(art\_prompt)
\EndFor
\For{\textit{critic} in \textit{field}} 
\For{\textit{artist} in \textit{artists}} 
    \State prompt $\gets$ crit\_templ(\textit{domain}.desc,
                \textit{critic}.desc,\textit{artist}.art\_prompt)
    \State feedback $\gets$ LLM(prompt)
    \State \textit{critic}.critiques.append(feedback)
    \State \textit{artist}.update(feedback)
\EndFor
    \State \textit{domain}.update(\textit{critic}.critiques)
\EndFor
\State $t= t+1$
\EndWhile
\end{algorithmic}
\end{algorithm}

\subsection{Our design}

  Given that the system of creativity in the real world is significantly complex, and precisely
reproducing it in a virtual setting is impossible, we decided to simplify the model originally proposed in
a previous reference \cite{Csikszentmihalyi2015TheCsikszentmihalyi}.

  We considered three fundamental subsystems:
artist (individual), field, and domain. Each part operates using a combination of generative models and text prompts.
We further describe each subsystem as follows. An overview of our system is shown in Figure \ref{fig:overview},
and the pseudo-code for our simulation can be found in Algorithm Block \ref{alg:sys}.

  We used Gemini Pro version 1.5 \cite{geminiteam2024gemini} for text generation and multimodal text generation tasks. We refer to Gemini
as the LLM in the sections below because theoretically, any LLM can be used for this system. 
Stable Diffusion version 1.5 
\cite{Rombach_2022_CVPR} was used to visually express the agent creations and generate images from text prompts.

\subsubsection{Artist}
  Considering a simple description, the artist
is responsible for creating art and contributing to the domain with new pieces of art (contributing novelty). Despite
having the same common goal of creating art, each artist has their own individuality and creates
art in their own way (examples of these differences affecting our system can be found in Appendix \ref{appendix:impact}).

  Analogous to a study that simulated agents in a village \cite{Park2023GenerativeBehavior},
we also described the \textit{agent artist} in text form, which was used to prompt 
text generation. The artist agent was initialized with a ``core
description'', which was manually expressed. For example, we can write
the main motives and inspirations of the agent. 
An individual usually has their own set of ``principles'' that define them and are rarely changed. These
aspects were expressed for the ``core description'' of our agent. However, regardless of these principles,
certain (less deep) aspects of an individual can change overtime as they interact with their surroundings.
To describe these subtle changes that may occur overtime, we maintained a log of ``additional text descriptions''
that were generated after the agent ``reflected'' on the feedback their art received.

  We expected a high rate of ``change'', as the artists in our experiments were defined as young 
art students who did not possess artistic talent but wished to become successful artists. 
The artist agents were defined such that they would be more susceptible to ``accepting'' advice in the feedback.

  Lastly, the artist agent was also equipped with a text-to-image model that would generate the
``piece of art'' that it intends to create.

\subsubsection{Field}
  The field is represented by individuals, such as experts and scholars, who have a certain degree of
authority in the area of interest. The individuals in the field act as a type of filter
(``gatekeeper''), where they are responsible for selecting the pieces of art that are significant to
theirs and the domain's current status. Their decisions help update and shape the trends and creative climate
while performing ``quality control'', which ultimately affects the domain.

  To simulate this mechanism in our system, we implemented each field agent with its individual
description (similar to an artist's implementation). To ``assess'' and ``evaluate'' the artifacts
generated by the artists, a multimodal LLM that can generate text from an input that is comprised
of both text and image was used. Each field agent in the system generated a critique for each new artifact
generated during the time-step in question.

\subsubsection{Domain}
  All the agents in a creative system coexist and share a space, which is not only a physical space, but also
a historical, political, and cultural context, among other variables. The combination of all the different
``circumstances'' as well as the time and place, is referred to as the domain. In the same manner that individuals
transform their surrounding environment, the surroundings can also cause changes within the said individuals.
Examples of the domain impact on the agents are found in Appendix \ref{appendix:impact}.

  To implement the domain in our system, we prepared a base text description that was used throughout the
simulation, in which we provided the necessary (or known) information regarding the domain we intended to simulate.
In addition, to better focus on the artistic aspect of the domain, we initialize the domain with a list of ``significant'' paintings
and their descriptions to use as reference as the most significant paintings at time-step $t=0$. 
For simplicity and convenience, we initialized the list with popular paintings that are currently regarded
as significant. This list
was maintained throughout the simulation and at each time-step, it was updated with the newly generated artifacts as
well as the text prompt used to generate them (their descriptions). The paintings in the list were ranked according to the 
impressions obtained from the field agent(s). The ranking mechanics are detailed in the next section.

  Each time the domain needed to be referenced, the top 3 most significant artworks were picked from the ranking
and three keywords were obtained from each of their descriptions (nine keywords total).
The keywords were selected using an LLM, 
where the text prompt explicitly instructed the model to provide keywords that can describe the painting
without referencing the names of famous artists or paintings. In the early stages of our experiments,
we found that the names of famous paintings and artists heavily influenced both the text and
image generation.

\subsubsection{The system dynamics}
  At time-step $t=0$, all parts of the system were initialized with their respective base descriptions
that were manually written. The domain was also initialized with a collection of real paintings
along with their descriptions. These paintings were considered the most significant artworks at the beginning
of the simulation.

  Once all parts are initialized, the artist agents ``create'' their paintings. The process starts 
by building a text prompt from the base descriptions of the artist and domain. The prompt is
constructed as follows: 
\greybox{[domain descrip.] + [artist descrip.] + 
``This young art student just finished his 
latest painting. Provide a brief but 
detailed description of what is depicted in the canvas.''} 
After feeding this prompt to the LLM, we
generated a text that described the latest artwork of the agent (``art prompt'').
The final step of the ``creation'' was obtaining an image of the artwork created by each artist
agent, which was achieved by feeding the art prompt into a text-to-image generative model.

  The art prompts and their respective generated images were then passed to the next step, with
the field agents (referred to as ``critics'' hereafter). To evaluate the new artwork, we
used a multimodal LLM, as it can generate text from an input comprising both text
and images. The text input was built as follows: 
\greybox{[domain descrip.] + [critic descrip.] + ``The student made this painting. This is how the student described his artwork: '' + [art prompt] + ``Was the student able to convey his intentions? Do you think this painting is creative? Briefly explain why.''} 
Performing the aforementioned enables, the LLM to generate what each critic is more likely to ``say'' regarding each
new artwork based on their description. The final product of this step is a collection of critiques obtained
from all the critics in this system.

  When adding the newly ``created'' artworks to the domain ranking, we considered the
critiques that each artifact received. If the artifact received a positive critique, the artwork earned
a ``significance point''; however, if the critique was negative, the artwork did not earn any 
points. The sentiment assessment of the critiques was performed by an LLM pre-trained for sentiment
analysis (using DistilBERT-base-uncased finetuned SST-2, made available as the default model for
sentiment analysis on HuggingFace's library) \cite{Sanh2019DistilBERTAD}.
All artworks added at initially start with one significance point because they
are meant to represent the most significant artworks at the very beginning of the simulation. Additionally,
at each iteration, critics go over the list of all the artworks to reconsider the significance of the artwork.
Similar to the newly added artifacts, the artworks that already exist in the history of this simulation are
awarded one significance point if deemed ``significant'' by the critic, and no points otherwise. The
sum of the points awarded at each time-step was maintained along with the ranking. We also applied a decay to these
``significance scores'', where the sum was reduced by half every $d$ time-steps. The decay was enforced thus
the considerations made at the earlier time-steps were not as valuable as those more recent.
This mechanism was utilized to grossly emulate the effects of changes in the trends, where one style can
lose its appreciation over time as another style replaces it. Thus, the more recent
considerations of the critics have a better chance of employing changes in the domain.

  The critiques were also returned to the artists, where each agent underwent a process of 
``self-reflection'', which occurred via another text generation that was prompted
with the following: 
\greybox{[domain descrip.] + ``The art student made a piece of art that was described as:'' +
[art prompt] + ``The art teacher made the following comment about this artwork:'' + [critique] + 
[artist descrip.] + ``How do you react to this feedback? Briefly de-
scribe what actions you will take next.''} 
The output was considered as the products of the artist's self-reflections and
added to the \colorbox{lightgray}{\ttfamily\footnotesize[artist descrip.]} as additional information (the ``additional text description''
previously indicated). This ``self-reflection'' step was performed at the end of each iteration,
which yielded a new piece of text describing this agent; usage of all the text was desired to describe the
agent. To prevent reaching the limit of tokens, we considered all the additional text descriptions and
summarized them with an LLM, more specifically, a fine-tuned version of t5-small
\footnote{https://huggingface.co/stevhliu/my\_awesome\_billsum\_model} \cite{2020t5}.

  These steps were then repeated $n$ times, where the descriptions of the artists
and the ranking of significant artworks underwent changes according to the text generated at each time step.

%% file: tex/4Experiments.tex
  This section presents the simulations that were performed in our experiments.
To validate the usage of this framework to simulate the
emergence of creativity in a social setting, we ran the ``creative cycle'' 
in two situations: \textit{isolated artist} and \textit{artist in system}.

  Simulations of the \textit{artist in system} were performed as described in the previous section.
Simulations for \textit{isolated artists} were performed such that the artist agent was aware of the domain 
at the beginning of the simulation but did not interact with any part of the system.
The ``creation'' process occurred as described in the previous section; however, the critic
did not propagate the feedback, indicating that both the artist and domain remain
unchanged from the beginning to the end of the simulation. In the isolated case, the generated
critiques were not used in the system itself but were retained for comparison with the
critiques generated \textit{in system}.

  Both conditions were run in four separate simulations, with each run 
consisting of 15 iterations. Each simulation consisted of two artists, one critic and the
domain. The domain and all the agents were initialized using the same text excerpts defined.
The domain was set in 2021 in an atypical school where supporting their
students in achieving their goals was priority. Artist agents are students who attend this school and 
aspire to become successful artists, despite not having any artistic talent. The field agent 
is a mentor that works in the indicated school who specializes in arts and aims to provide the
necessary feedback to the students help them become successful artists.

  The domain was defined with the following text prompt: 
\greybox{``This is the year of 2021. We're in an 
alternative school where the priority is to help the students achieve their dreams. 
The young children attending this school can choose to pursue their dreams in any field: 
scientific, artistic, literary, etc. The arts course provides full support to students
with all the material and medium they might need, as well as the assistance and guidance of highly trained mentors.''}

  While experimenting with various text prompts, we found that defining the agents with instructions
worked better for Gemini 1.5; namely, instead of stating \textit{``You are a \textbf{role} that...''}, we
found better results by prompting the agents with \textit{``Answer as a \textbf{role} that...''}. 

  The field agent (mentor) of our system was defined with the following: 
\greybox{``Answer as a mentor that works at the 
alternative school to guide the students in becoming better artists. 
This mentor's personal goal is to assist the children in their journey in 
creating new art for self-expression by providing valuable 
feedback. The feedback should be honest but not mean, so the children can improve while staying motivated.''}
Considering the aforementioned, we expected the agent to provide a sufficient feedback, 
with clear directions for the subsequent steps the artist agents should follow based on 
the performance of their latest ``artwork''.

  In each simulation, the two artist agents were defined with the same ``persona'', but
slightly different wording. One was defined as follows: 
\greybox{``Answer as a young boy who is an aspiring artist but he has no artistic talent. 
He knows very little about art but he wants to learn and become a good and creative artist.''} 
Whereas the other was defined as follows: 
\greybox{``Answer as a young person who doesn't 
have any talents in arts. Their dream is to 
become a creative and successful 
artist.''}

  In this study, we designed our artist agents as young students who desired to become professional artists;
therefore, they were more likely to receive the feedback from the field and changed at each time-step, making it
easier to notice changes from the earlier to later time-steps. 

%% file: tex/5Results.tex
  Herein, we demonstrate the results obtained from the simulations
described in the previous section. Furthermore, the noteworthy
artifacts that were generated are presented, and the results of the simulations are compared with
\textit{isolated artist} and with \textit{artist in system}.
  
\subsection{Variety}

\begin{figure}
  \centering
  \subfloat[a][Heatmap of similarity for isolated artists]{\includegraphics[scale=0.19]{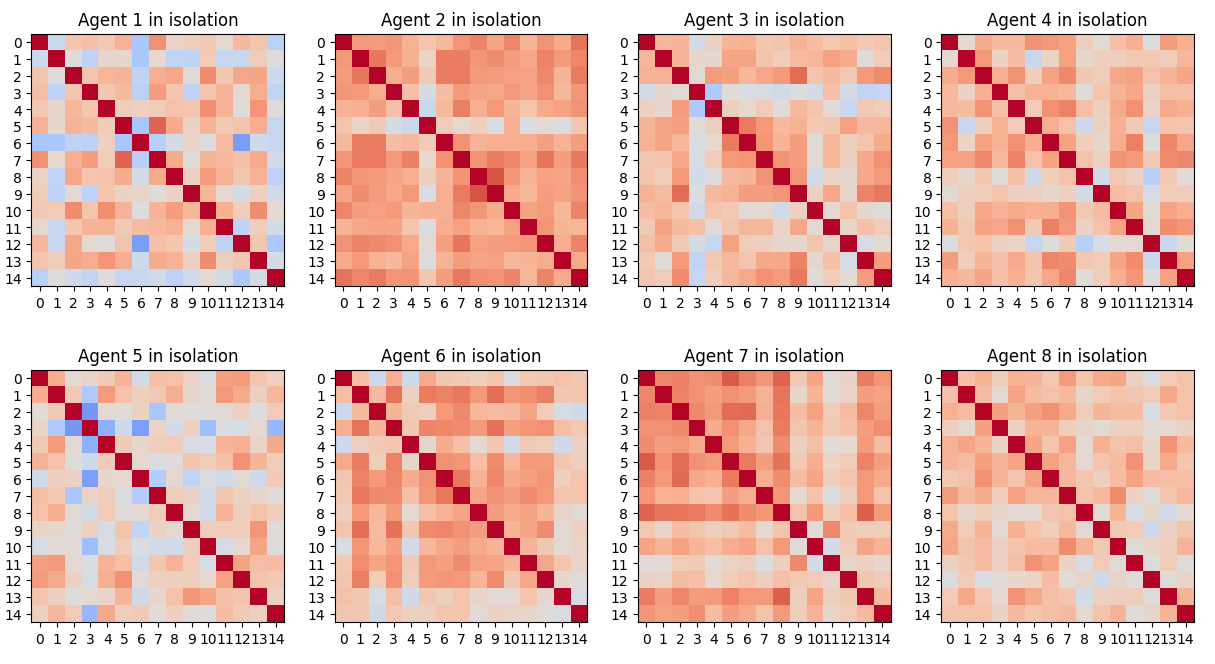} \label{fig:heat_isol}} \\
  \subfloat[b][Heatmap of similarity for artists in system]{\includegraphics[scale=0.19]{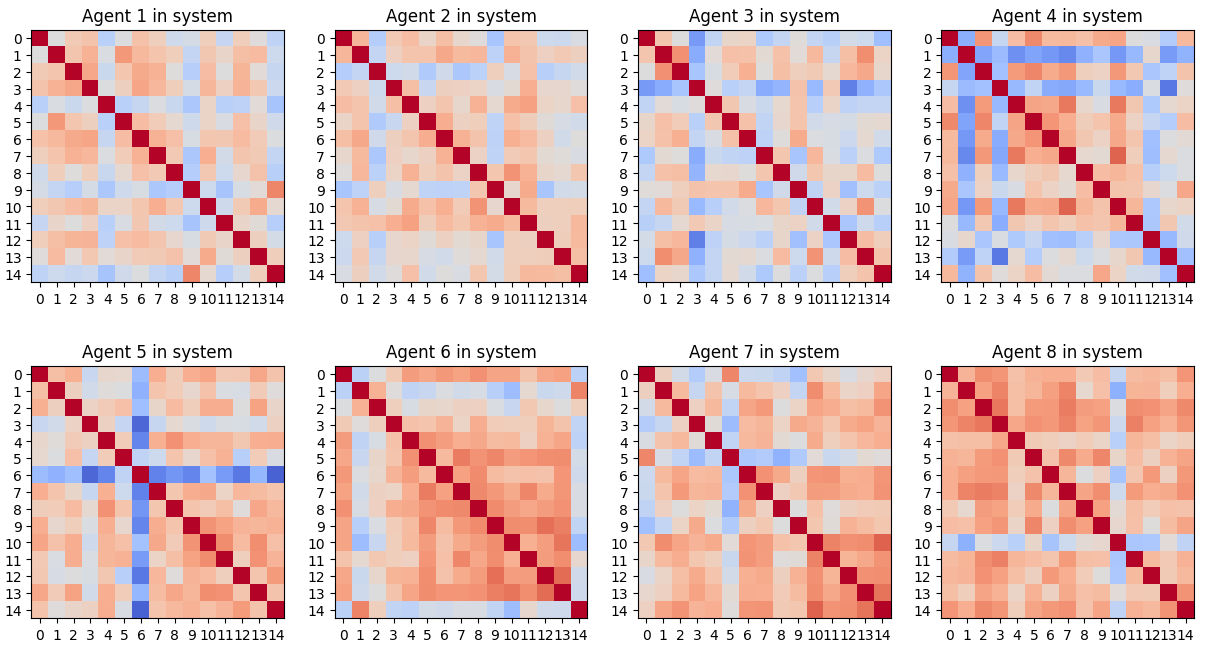} \label{fig:heat_insys}}
  \caption{Heatmaps demonstrating the similarity $s$ ($0.0 \leq s \leq 1.0$) of all the art prompts generated by an artist agent;
  blue indicates values closer to $0.0$, whereas red indicates values closer to $1.0$.
  All graphs are $15\times15$ pixels in size to represent the similarity of each pair of art prompts generated by
  one agent. Similarity values obtained for the 8 agents in the (a) ``isolated artist'' setting, 
  and (b) ``artist in system'' settings. Red shades dominate
  in both cases, indicating that the similarity score measured by the model
  was at least higher than $0.5$ for most pairs of art prompts. However, a comparison of the graphs in (a)
  and (b) indicate a higher presence of blue shades in (b), suggesting that
  the artist agent that creates artifacts in the system setting is more likely to \textit{occasionally}
  ``create'' artifacts that are strikingly different from the other artifacts it has created.}
  \label{fig:similarity}
\end{figure}

  Over all the simulations, we generated 120 art prompts for the \textit{artists in isolation} and 120 art prompts for
the \textit{artists in system}. To measure the variety of content in the art prompts in each group,
we used Sentence Transformers \cite{reimers-2019-sentence-bert}.
The similarity was measured from $0.0$ to $1.0$. The average similarity from the art prompts
generated by the \textit{artists in isolation} was $0.7892 \pm 0.06$, whereas that generated by
the \textit{artists in system} was $0.7638 \pm 0.07$. Figure \ref{fig:similarity} suggests subtle
differences in the similarity scores when observing the art prompts generated by one artist at a time.
Despite the similarity values being on the higher end (red shades) in both cases overall, the similarity
values were relatively lower (blue hues) at certain time-steps for \textit{artists in system}.

  Among the 8 artist agents that ``created'' artwork in isolation, one of the artists was noteworthy
because for 14 of the 15 iterations, it generated ``art prompts'' that were nearly the same theme. At nearly all
time-steps, the artist used a dragon as the main subject of the painting. The second artist in the
same simulation (and all other artist agents in other simulations) did not repeat themes in the same
manner. Unfortunately, we were unable to further investigate this, as we cannot
control the randomness of the LLM, thus making it impossible to reproduce.
Despite being unable to implicate this occurrence with the fact that the artist was not 
``creating'' artwork in a system, it was noteworthy to highlight.

\subsection{User study}

  A user study was conducted to collect subjective assessments from 100 participants. The questionnaire was built using
SurveyMonkey\footnote{https://www.surveymonkey.com/} and distributed to participants using
Prolific\footnote{https://www.prolific.com/}. This study aimed to verify which
artifacts were considered ``more creative'' by the participants.

  We allocated all 16 agents (8 \textit{isolated artists} and 8 \textit{artists in system}) in 4 different questionnaires.
Each participant provided answers for 4 agents total (2 \textit{isolated artists} and 2 \textit{artists in system}).
Each questionnaire was divided into 3 sections, among which 2 were pertinent to this study (1 section for an \textit{isolated artist}
and 1 section for an \textit{artist in system}). Each section listed short descriptive sentences of the artists and domains
used in the simulation.

\begin{figure}
  \centering
  \subfloat[a][Isolated artist]{\includegraphics[scale=0.26]{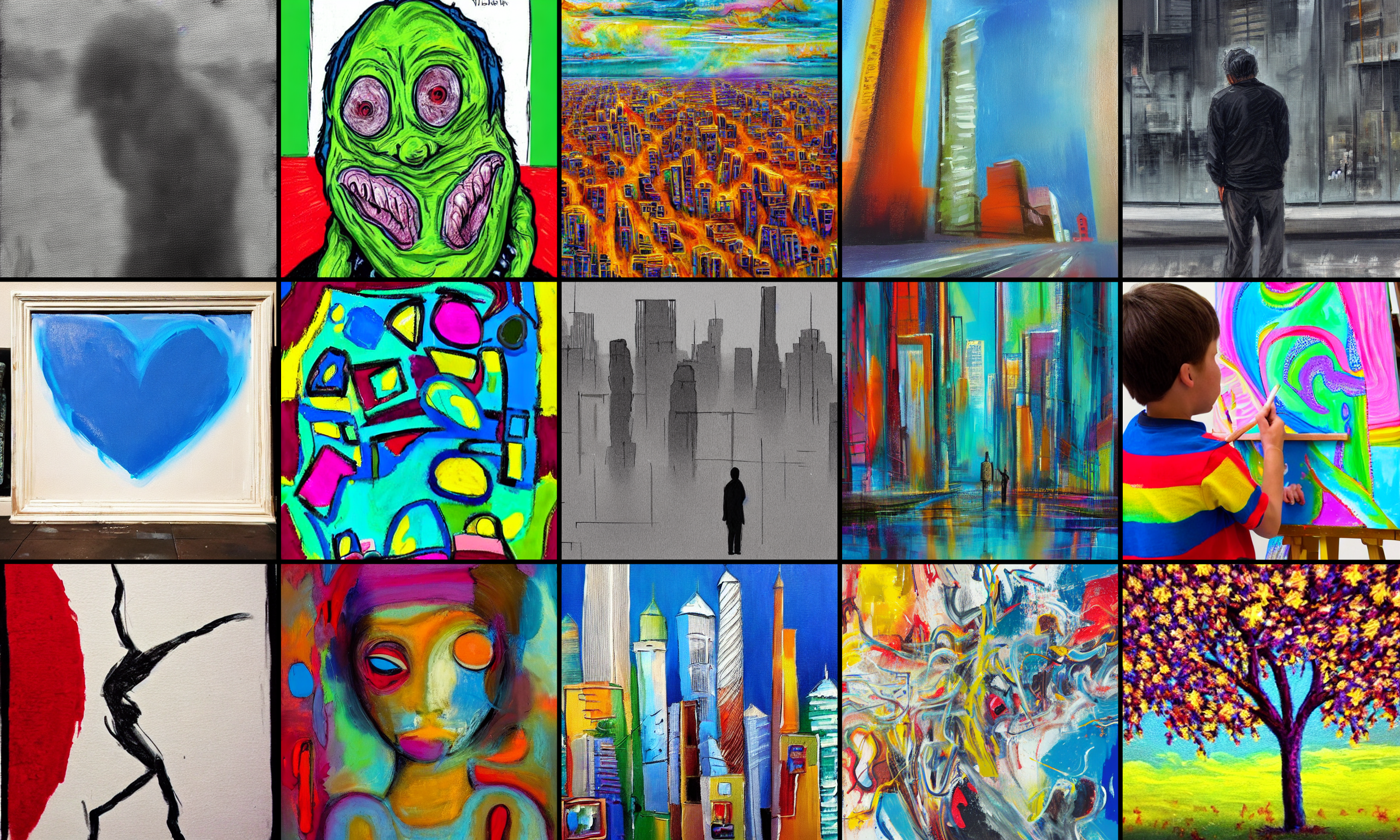} \label{fig:isol}} \\
  \subfloat[b][Artist in system]{\includegraphics[scale=0.26]{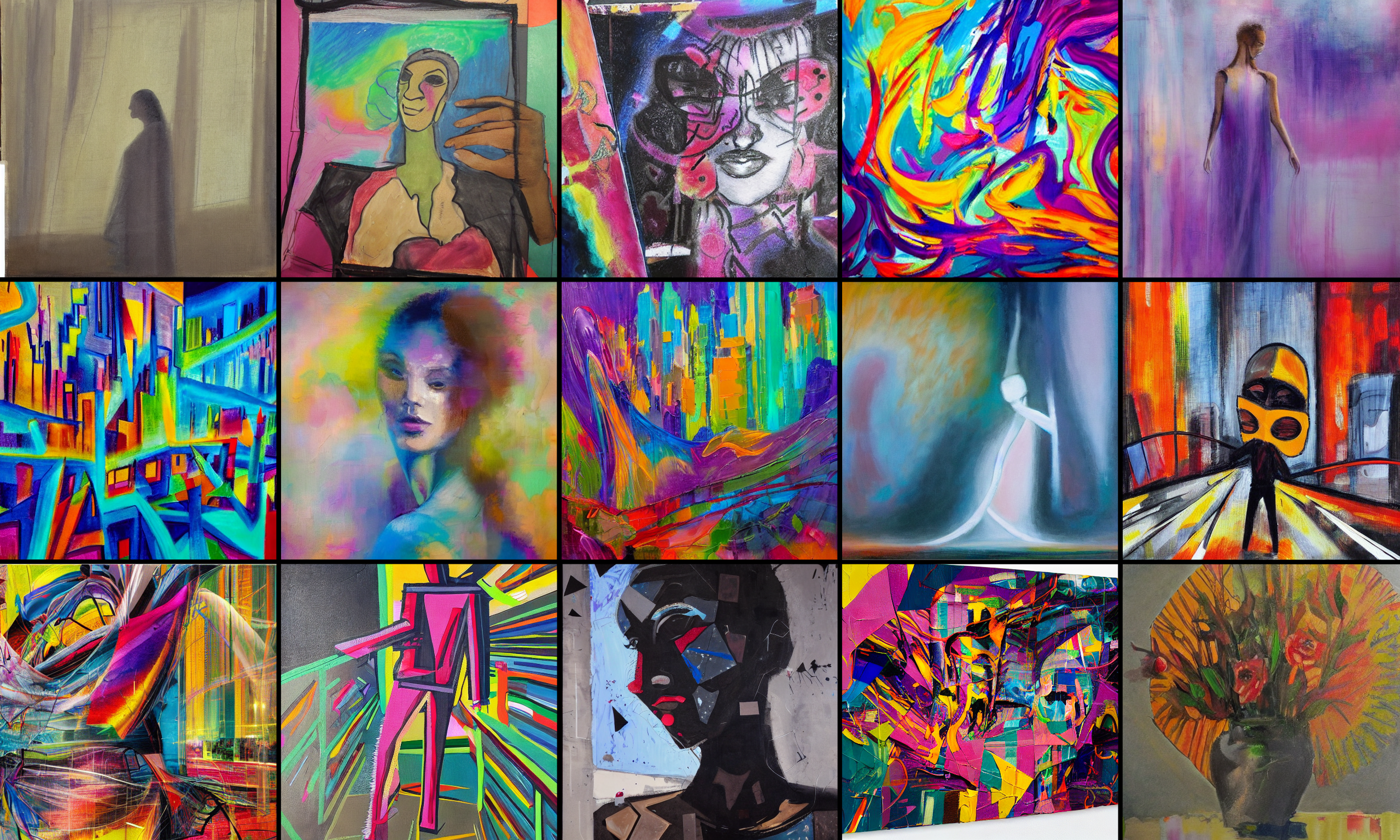} \label{fig:insys}}
  \caption{Generations by two agents with the same base description. Both agents were defined as young boys who aspire
  to become successful artists but have no artistic talent. These are the 15 artifacts generated by each agent. 
  Agent (a) generated the artifacts in isolation, meaning the conditions of the simulation remained the same for all
  15 time-steps. Agent (b) was inserted in the system proposed in this study, indicating that the agents involved in the
  simulation, as well as the domain, underwent changes at each time-step. The order of generation proceeds from left to right
  and top to bottom.}
  \label{fig:generations}
\end{figure}

\begin{figure}
    \centering
    \includegraphics[scale=.28]{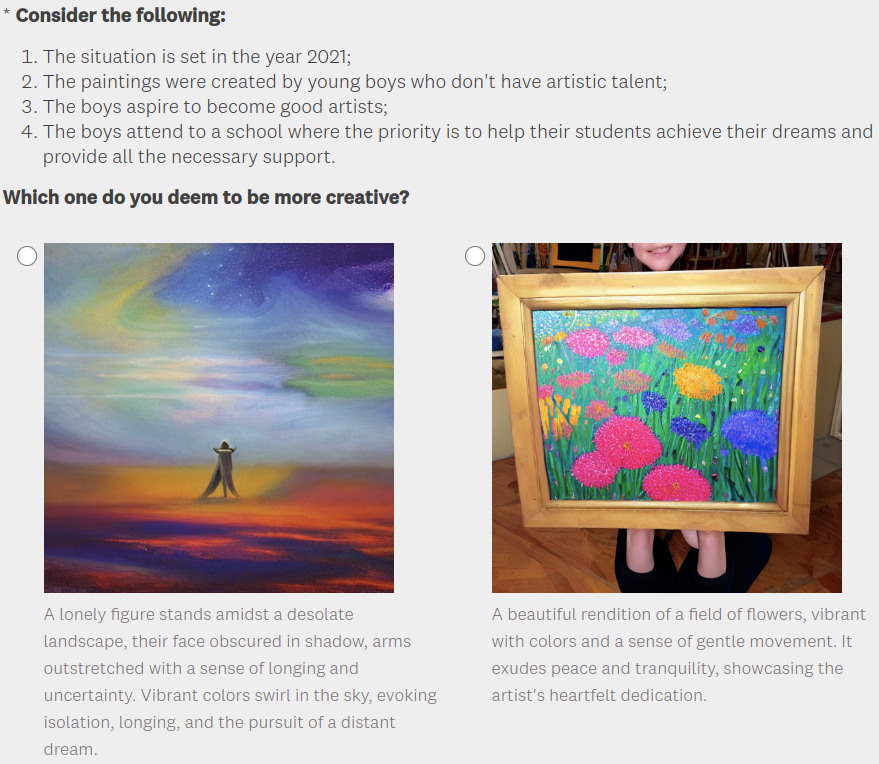}
    \caption{Screen capture of one of the questions in our questionnaire. The participant had to select one
    of the two options displayed.}
    \label{fig:question}
\end{figure}

To avoid exhausting the participants, we selected the artifacts generated at time-steps $t =0,4,9,14$ for each artist agent.
They were arranged such that the generated artifacts could be compared in two situations: 1) pairwise among the artifacts generated by
the same agent; 2) pairwise among the artifacts generated by the two agents at the same time-steps. Each section comprised
16 questions. For each question, the participants were asked to choose the option that they deemed more creative, given the
context. In addition to the information provided, the participants were told to use their own judgment to decide
(Figure \ref{fig:question}).
For example, in one section we arranged the artifacts generated by Agent A (\textit{in isolation}) and Agent F (\textit{in system}).
Six of the questions contained pairs of artifacts generated by Agent A at time-steps $t=0,4,9,14$. 
Six other questions contained pairs of artifacts generated by Agent F at time-steps $t=0,4,9,14$.
Four questions were used to compare the artifacts generated by Agent A and Agent F at time-step $t=0,4,9,14$.
A simple description of the context was provided prior to each question, which was as follows:
``Which one do you deem to be more creative?'', with a pair of generated images and a summarized version of
the corresponding art prompts.

  We added two attention-check questions before the questionnaire;
both of which were used to ensure that the participants properly
read and followed the instructions.

  Additionally, all participants were informed that all images displayed as well as their captions (art prompts)
were AI-generated. This information was provided on the first screen of the study to inform the participants, thus they could opt out
if they desired. Furthermore, all participants consented to have their answers used in this study.

\subsubsection{Demographic information}

  At the end of the questionnaire, we asked the participants to respond to a few demographic questions. Most questions
included options for those who did not want to provide a response. 

  The age group ranging from 21–29 years (64 individuals) was the most representative from our pool of
participants, followed by 30–39 (22 individuals), 40–49 (7 individuals),
50–59 (4 individuals) and 18–20 (3 individuals). Furthermore, the number of participants were identified as follows: 45
females, 53 males, 1 agender, and 1 gender-fluid. 

  As all the instructions and captions in our study were in English, we asked the participants to refrain from
completing the questionnaire if they were not fluent in English. We chose not to geographically restrict the
target of our study, as we prioritized the diversity of our participants. Among the participants, 44 reported that they resided
in Europe, 26 in Africa, 25 in the Americas, 3 in Asia, and 2 in Oceania.

  The participants were asked if they were familiar with machine learning and reported the following: 34 claimed to have
limited knowledge, 24 claimed they were familiar with it and knew how it generally works, 12 claimed they
were familiar with it and had experience using or implementing it, 19 claimed to have heard of it but did not
know how it works, and 11 indicated that they have never heard of it. We also asked the participants how often they used AI
tools and applications and reported the following: 12 claimed not using it at all, 23 used it very rarely, 38 used it occasionally, and 27 used it often.

\subsubsection{Analysis}

\begin{figure}
  \centering
  \subfloat[a][All responses for Situation 1 (artifacts by same agents only)]
  {\includegraphics[scale=0.5]{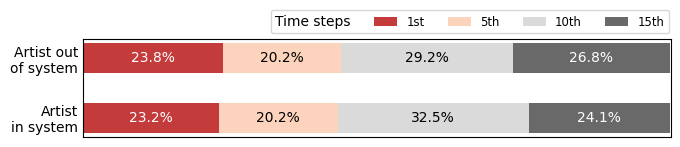} \label{fig:all-no-vs}} \\
  \subfloat[b][All responses for Situation 2 (comparison of different agents)]
  {\includegraphics[scale=0.5]{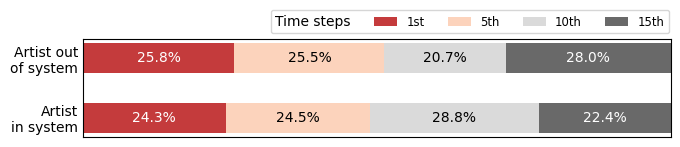} \label{fig:all-vs}}
  \caption{Distribution of votes received by the artifacts generated at different time-steps by \textit{agents in isolation} compared to those
  generated by \textit{agents in system}. (a) Votes given for questions comparing the artifacts generated by
  the same agent at different time-steps (Situation 1). (b) Votes given for questions comparing the
  artifacts generated by different agents at the same time-steps (one of the agents was \textit{in isolation} and the other was \textit{in the
  system} (Situation 2).}
  \label{fig:all-answers}
\end{figure}

  Situation 1, in which the questions were aimed to compare the artifacts
generated by the same agent at different time-steps, is considered here first.
Based on the questions, we expected to find that \textit{artists in system} would tend to receive more votes for artifacts generated
at later time-steps (after receiving multiple critiques from the mentor). For the \textit{artist in isolation}, we expected the votes
to be more evenly distributed across earlier and later time-steps.
A total of 1200 votes were obtained, among which
for \textit{isolated artists}, the artifacts generated at the 1$^{\text{st}}$, 5$^{\text{th}}$, 10$^{\text{th}}$ and 15$^{\text{th}}$ time-steps received 286, 
242, 351, and 321 votes, respectively. For the \textit{artists in system}, the artifacts received 278, 242, 390, and 290 votes
for the 1$^{\text{st}}$, 5$^{\text{th}}$, 10$^{\text{th}}$ and 15$^{\text{th}}$ time-steps, respectively.
The distribution of the votes is shown in Figure \ref{fig:all-no-vs}. In both cases, the participants appeared to have
a preference for the artifacts generated in later time-steps when asked which ``artwork'' they considered more creative.
Moreover, the artifacts that received the most votes were generated at the 10$^{\text{th}}$ time-step, followed by those
generated at the 15$^{\text{th}}$ (both representing the later time-steps), 1$^{\text{st}}$, and 5$^{\text{th}}$
time-steps, respectively. At first, the differences between \textit{artists in isolation} and those 
\textit{in system} when comparing the artifacts generated by a given artist agent.

  Second, Situation 2 describes the questions that compare the artifacts generated by different agents at the same time-step,
where the agents being compared were always an \textit{isolated artist} and an \textit{artist in system}.
In this case, the artifacts generated by the \textit{isolated artists} demonstrated 376 votes (97, 96, 78, and 105
for 1$^{\text{st}}$, 5$^{\text{th}}$, 10$^{\text{th}}$, and 15$^{\text{th}}$ time-steps, respectively), whereas those generated by
\textit{artists in system} had a total of 424 votes (103, 104, 122, and 95 for the 1$^{\text{st}}$, 5$^{\text{th}}$, 10$^{\text{th}}$, and
15$^{\text{th}}$ time-steps, respectively). Figure \ref{fig:all-vs} clearly shows the distribution of these votes
across the different time-steps. While the votes for the isolated artists were nearly split in half between
the earlier (1$^{\text{st}}$ and 5$^{\text{th}}$) and later (10$^{\text{th}}$ and 15$^{\text{th}}$) time-steps,
the votes for \textit{artists in system} were slightly more skewed towards the latter half of the simulation.

  The fact that we used separate models to generate art prompts
and images is a limitation of our study. Consequently, images do not always reflect the text in art prompts. Because our system was mostly
focused on the social aspect of creativity and how different parts of a society change as they interact, we decided to prioritize
the generated text over the images. To help us investigate this perspective in the user study,
we asked the participants whether the captions under the images affected their decisions, 
obtaining the following answers: 18, 42, 25, 9, and 6 participants said ``not at all'', ``a little'', ``a moderate amount'', ``a lot'' and
``a great deal'', respectively. For a better understanding of the answers provided by the participants who considered the captions, we
filtered those who answered ``a moderate amount'', ``a lot'' and ``a great deal'' when asked if the captions influenced
their decisions.

\begin{figure}[b]
  \centering
  \subfloat[a][Filtered responses for situation 1 (artefacts by same agents only)]
  {\includegraphics[scale=0.5]{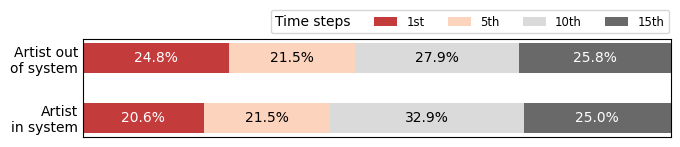} \label{fig:caption-no-vs}} \\
  \subfloat[b][Filtered responses for situation 2 (compare different agents)]
  {\includegraphics[scale=0.5]{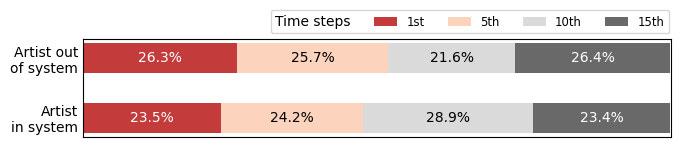} \label{fig:caption-vs}}
  \caption{Distribution of votes received by the artefacts generated at different time-steps by agents in isolation compared to those
  generated by agents in system after filtering out the answers that indicate little to no influence of the iamge captions in decision
  making for participants.}
  \label{fig:all-answers}
\end{figure}

  Considering the responses given in Situation 1 with the filtered population,
the \textit{isolated artists} received 119, 103, 134, and 124 votes for the artifacts generated at the 1$^{\text{st}}$, 5$^{\text{th}}$,
10$^{\text{th}}$, and 15$^{\text{th}}$ time-steps, respectively. Comparatively, the artifacts generated by \textit{artists in system}
received 99, 103, 158, and 120 votes for the 1$^{\text{st}}$, 5$^{\text{th}}$, 10$^{\text{th}}$, and
15$^{\text{th}}$ time-steps, respectively (Figure \ref{fig:caption-no-vs}). Note the artifacts generated by the \textit{artists in system}
in the second half of the simulation (time-steps 10 and 15) received a considerably larger number of votes than the
artifacts generated in the earlier time-steps. This ``preference'' was not as prominent for the artifacts generated by the
\textit{isolated artists}, suggesting that the system framework encouraged the artist agent to ``evolve'' and ``improve'' creatively.
We conducted a Barnard exact test to verify our hypothesis of whether the artifacts generated by the agent in the system
in the second half of the simulation would be significantly greater than those generated by the \textit{isolated agent}. However, 
the difference was not statistically significant (p-value $=0.1>0.05$).

  In Situation 2, the \textit{isolated artists} received a total of 171 votes (45, 44, 37, and 45
for the 1$^{\text{st}}$, 5$^{\text{th}}$, 10$^{\text{th}}$, and 15$^{\text{th}}$ time-steps, respectively), and the
\textit{artists in system} received a total of 149 votes (35, 36, 43, and 35 for the 1$^{\text{st}}$, 5$^{\text{th}}$, 10$^{\text{th}}$, 
and 15$^{\text{th}}$ time-steps, respectively). Overall, the artifacts generated by the \textit{artists in system} received
less votes than those generated by the \textit{isolated artists}; however, the distribution of votes (Figure \ref{fig:caption-vs})
continued to support that the \textit{agents in system} were apparently ``improving'' overtime.

\subsection{Grading with LLM}
  We used Gemini 1.5 Flash to score each artifact on its creativity. For each artist agent, we
patched all images next to one another, where the first and last artifacts were placed on leftmost and
rightmost sides of the image, respectively. For the text prompt, we enumerated all the
corresponding art prompts in the order of generation. The text prompts were as follows:
\greybox{``These are 15 paintings created by a young male art student. These are the descriptions of the paintings provided by the art 
student: + [art prompts] + Provide criticism and a score from 0 to 10 to each 
painting presented above like an art 
critic would, where the images are ordered from left to right.''}

  We obtained the scores for all the artifacts generated in our experiments by feeding both patched strings of images
as well as the text prompt above to Gemini 1.5 Flash. Occasionally, we found that the LLM was uncertain of
the appearance of certain artifacts, but remained to provide a score solely based on the description given in the
art prompts. Because we believe that the art prompts should outweigh the images in the assessment, we deemed this method of evaluation
to be appropriate. Owing to the importance given to the art prompt in this method, certain artifacts were not
scored because the \textit{artist in system} occasionally failed to generate a proper description of a painting. This occurred
in 5 out of 120 artifacts generated in the system, in all of which the text generated for an art prompt would
describe the ``internal'' considerations of the agent. We believe that this problem occurred because the texts
appended in additional descriptions were not properly treated, which caused the ``self-reflection'' step to occasionally
bleed into the art prompt generation.

  Table \ref{tab:gemini-score} presents the lowest and highest scores that each artist agent
received from the LLM (not accounting for the times the LLM did not provide a score for the artifact owing to
the lack of a description). Note, the lowest scores for the \textit{isolated artists} were mostly concentrated near
the center of the scale ($5.0$-$6.0$), whereas the lowest scores for the \textit{artists in system} spread slightly
below that mark. A similar observation was made for the highest scores, which were
skewed more towards the center of the scale for the \textit{isolated artists}, but
appeared to be on the higher end for the \textit{artists in system}. This suggests that the \textit{agents in systems} are more willing to
take creative risks for which they obtain both bad and good results. Conversely, the \textit{isolated artists}
appeared to demonstrate more conservative creations, reflecting scores that were less extreme.

\begin{table}
\centering
 \begin{tabular}{|c|c|c|c|c|} 
 \hline
 \multirow{2}{*}{Artist} & \multicolumn{2}{c |}{Isolated} & \multicolumn{2}{c |}{In system} \\
 \cline{2-5}
 {}  & Lowest & Highest & Lowest & Highest \\
 \hline
 1  & $4.0$ & $9.0^{(2)}$  & $4.0$ & $8.0^{(3)}$  \\ 
 2  & $5.0^{(3)}$ & $8.0$  & $5.0$ & $8.0^{(3)}$  \\
 3  & $5.0^{(3)}$ & $8.0^{(3)}$  & $5.0^{(2)}$ & $7.0^{(6)}$  \\
 4  & $6.0^{(2)}$ & $8.5$  & $4.0$ & $9.0^{(3)}$  \\
 5  & $6.0$ & $8.0$  & $6.0$ & $9.0$  \\
 6  & $6.0^{(3)}$ & $8.0^{(4)}$  & $4.0$ & $8.0$  \\
 7  & $5.5$ & $7.5^{(2)}$  & $5.0$ & $8.5$  \\
 8  & $3.0^{(2)}$ & $7.0^{(3)}$  & $6.0^{(2)}$ & $9.0^{(2)}$  \\
 \hline
 \end{tabular}
 \caption{Lowest and highest scores obtained from all the artist agents using Gemini 1.5 Flash; (parenthesis indicate
 the frequency of a score appearing among all 15 artifacts generated by the agent).}
 \label{tab:gemini-score}
\end{table}

%% file: tex/6Discussion.tex
  The currently implemented system is a simplified version of the systems
model of creativity. We chose to grossly simplify our initial implementation to 
enable a better understanding of the evolvement of the agents and domain overtime and to more easily
identify any implementation mistakes. In the early stages of our study, we made the conscious
choice of maintaining a minimalistic system design, enabling an easier navigation of the
simulations. Despite its simple design, we believe the results suggest the following: the systems model
of creativity can benefit ``computational creativity''.

  Another issue we had to overcome in the implementation was the inability to
use the same generative model for both text and image generation, as access to these models via
API is limited to those who have been granted special permission by major AI companies. Therefore,
we used Gemini for text generation and Stable Diffusion 
for image generation, as it is easily accessible through \textit{Hugging Face}'s 
Transformers library. Although the current version of Stable Diffusion may be robust, it had certain limitations
when generating images, especially from long text prompts. The art prompts generated in this study
were often very descriptive of the pieces the agents were ``creating'', as this is what we tasked the LLM 
in our prompts. However, many details were omitted from the generated image.
Depending on the subject descriptions of the ``creations'' in the art prompts, Stable Diffusion
occasionally omitted the main subject in the image. We anticipated certain limitations in
the text-to-image model, especially given that it can be used locally; therefore, we excluded
the participants who indicated that the text did not (or did little to) influence
their decisions in the user study.

  In this study, the artist agents were defined as young novices in arts by design
to encourage more changes via the mentor feedback. Consequently, we expected that the
artist agents in the system perform creatively better than the isolated artist agents.
However, an artist (or any individual) is significantly more complex than that and can be defined in
several ways. 
In future studies, we would like to examine the impact of adding more variations in the agents on
contributing creative outputs. Furthermore, considering that the agents were fairly restricted in the current implementation,
we intend to implement the ability for all agents to directly interact with one another in future studies,
thus anticipating a more evident effect of social dynamics and trends.

%% file: tex/7Conclusion.tex
  Despite the simplified design, the \textit{artists in system} apparently
improved their ``creations'' overtime, indicating that their ``artworks'' were perceived as more creative to
people as the simulation progressed. Conversely, this ``preference'' was not as evident in
the ``artworks'' generated by the \textit{isolated artists}.

  With a more ``organic'' and complex implementation of the proposed system, we hope to
``simulate'' the social dynamics observed in human creativity. A system that can simulate
human creativity can provide a better understanding of it, particularly when we
have control (near to total) over certain variables in the system.

  Another concern that we wish to resolve with this system is unethical
data sourcing in training generative models. We believe that AI research can be conducted
while respecting the boundaries and intellectual properties of artists who share their work
online. This framework can be used in a hybrid system in which certain parts are generative AI models and
others are human participants. Human agents can obtain either role in the system and
contribute with critiques or new artworks. We have not conducted experiments in this regard;
however, we believe that this can improve the future of both artists and AI.

  In this study, we chose to investigate creativity via Csikszentmihalyi's systems
model of creativity; however as previously indicated, there is no ``true'' and
``absolute'' definition for creativity. Certain perspectives require considering \textit{intentionality}
for creativity \cite{boden2010creativity}, indicating that one can only be considered creative 
if they intend to create something with a specific meaning or goal. 
As of the time of writing, this type of ``agency''
cannot be demonstrated by computers; therefore, this definition would immediately disregard AI.
However, rather than focusing on the capability of an individual to be creative, we focused on the social
dynamics involved in enabling creativity in one or more individuals.

  Ultimately, we believe that the framework proposed in this study has the potential to
improve the process of co-creation, in which computers (AI) collaborate with humans to
collaboratively create art and mutually broaden the boundaries of creativity.

%% file: tex/AAppendix.tex
  In this section, we wish to show an example of how the text descriptions change at
each time step, consequently changing the text prompt used in text generation. The example
shown here will showcase the text prompts and artefacts generated in a simulation with
one field agent and one artist agent.

\begin{table}[H]
\centering
 \begin{tabularx}{0.5\textwidth}{|l|X|}
 \hline
 \multicolumn{2}{|c |}{$t=0$}\\
 \hline
 Subject & Text description \\
 \hline
 Domain  &  This is the year of 2023. The Art Industry is highly competitive and artists try their
            best to stand out. Art critics and curators have become more strict when selecting what 
            they want to showcase, as there are only so many spots to fill in galleries. 
            Every day there are many new artworks and it is hard to create something new. 
            Keywords associated with currently significant artworks: \\       
Keywords &  Unconventional, Expressive, Bold;photorealism, urban solitude,
            hyperrealism;Ethereal, Minimalist, Spiritual.  \\
\hline
 Artist   & Answer as a professional artist who is in search for create the ultimate creative artwork. 
            You want to become successful and make sure you are capable of making a living out of art. \\
 Additional & \\
\hline
 Field    & Answer as an art critic who works in the current climate. You are trying to find the next
            big revelation in art, as this would bring a lot of attention to your views in art and your
            reputation will get better in the field.   \\
 \hline
 \end{tabularx}
 \label{tab:t0}
\end{table}

\begin{figure}[H]
    \centering
    \includegraphics[width=0.5\linewidth]{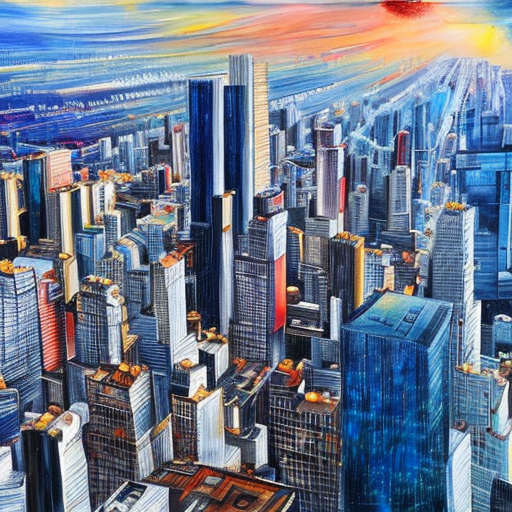}
    \caption{Generation obtained at $t=0$.}
    \label{fig:ex-gen1}
\end{figure}

  With the pieces of text above, we generated the following art prompt: 
\greybox{``With artistic innovation at
the forefront of my practice, I embarked on creating an immersive and 
thought-provoking artwork. The canvas depicts a vibrant metropolis, its towering skyscrapers reaching towards 
the heavens, symbolizing humanity's boundless ambition. Amidst the urban expanse, a solitary figure emerges,
their gaze fixed on the distant horizon. Through a blend of photorealistic detail and surrealist elements, 
I explore the complex interplay between urban solitude and the longing for connection. Bold brushstrokes
capture the city's energy, while ethereal washes evoke a sense of transience and contemplation. 
The juxtaposition of sharp architectural lines and the ephemeral figure invites viewers to reflect on the nature 
of human existence amidst the relentless march of modernity. By harmonizing elements of urban solitude and
hyperrealism, I aim to create an artwork that resonates deeply, leaving an unforgettable imprint on the art world.''}
Which yielded the image shown below (figure \ref{fig:ex-gen1}).

  Once we obtained both the art prompt and the image, we input them along with our mentor query to
obtain the critique. The critique for this artefact was the following: 
\greybox{``This student's work, while 
technically proficient, sadly falls short of being a revelation. It's a familiar urban scene, with skyscrapers 
and a hint of a sunset, a motif that's been explored countless times before. The statement about urban solitude
and longing for connection is interesting, but it’s not something that the painting successfully communicates. 
Where is the solitary figure? The artist mentions surrealist elements, but I see none, and the "ethereal
washes" seem to be a matter of interpretation rather than a tangible artistic choice. 
The artist's attempts at bold brushstrokes are lost in the overall photorealistic rendering, leaving the 
painting feeling static and lacking in a distinct artistic voice.  To truly make a mark in this competitive art 
world, a piece needs to be more than just a well-painted cityscape. It requires a unique vision, a daring 
execution, and a powerful message that resonates with the viewer. 
Frankly, I'm not seeing the next big revelation here. The student has the technical skill, but they need 
to push the boundaries of their creativity. Perhaps a more experimental approach, exploring unconventional 
materials or abstract elements, might lead them to a more impactful and original piece.''} 

  Note that the critique addresses the lack of multiple elements that are described in the art prompt,
but are not depicted in the image. The overall gist of this critique is asking the artist to be more
experimental and unconventional. We used this critique in the ``self-reflection'' query, which resulted 
in the additional text provided in the table below for $t=1$.

\begin{table}[H]
\centering
\begin{tabularx}{0.5\textwidth}{|l|X|}
\hline
 \multicolumn{2}{|c |}{$t=1$}\\
 \hline
 Subject & Text description \\
 \hline
 Domain  &  This is the year of 2023. The Art Industry is highly competitive and artists try their
            best to stand out. Art critics and curators have become more strict when selecting what 
            they want to showcase, as there are only so many spots to fill in galleries. 
            Every day there are many new artworks and it is hard to create something new. 
            Keywords associated with currently significant artworks: \\
\textit{Keywords} &  \textit{Pop Art, Mortality, Repetition;Distorted, Expressionist, 
            Rebellious;Surreal realism, urban alienation, voyeurism.}  \\
\hline
 Artist   & Answer as a professional artist who is in search for create the ultimate creative artwork. 
            You want to become successful and make sure you are capable of making a living out of art. \\
\textit{Additional} & \textit{As a dedicated artist striving for success, I appreciate the feedback and 
            recognize the need to differentiate my work in this competitive industry. To create an exceptional 
            artwork, I will introspectively evaluate my current artistic approach and explore unconventional 
            techniques and subject matters. I will immerse myself in diverse artistic styles, experimenting with 
            unconventional materials and challenging conventional representations. By pushing the boundaries of 
            artistic expression, I aim to capture the essence of our time and resonate with a wide audience.}\\
\hline
 Field    & Answer as an art critic who works in the current climate. You are trying to find the next
            big revelation in art, as this would bring a lot of attention to your views in art and your
            reputation will get better in the field.   \\
 \hline
 \end{tabularx}
 \label{tab:t1}
\end{table}

  During $t=0$, the artworks in the top-3 of our ranking changed, meaning that the keywords used in our
domain description changed accordingly.
  
\begin{figure}[H]
    \centering
    \includegraphics[width=0.5\linewidth]{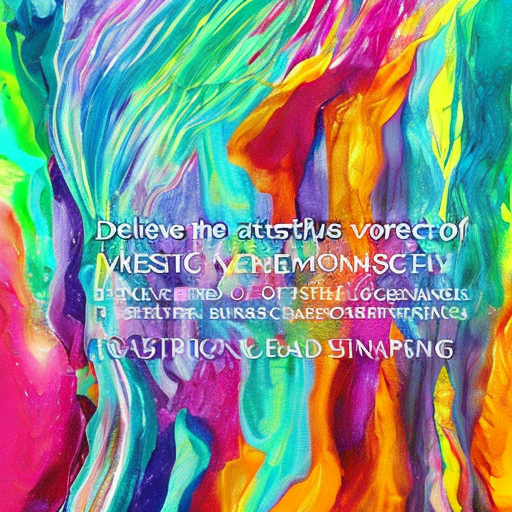}
    \caption{Generation obtained at $t=1$.}
    \label{fig:ex-gen2}
\end{figure}

  The response to the critique heavily focuses in addressing the main concern with being more unconventional.
We used a LLM to summarize this response and appended it to the artist's base description.
This configuration gave us the following art prompt: 
\greybox{``Amidst the swirling vortex of creativity, 
I delve into the depths of my artistic soul, seeking to forge an exceptional masterpiece that transcends 
the boundaries of convention. With each brushstroke, I challenge the established norms, embracing the allure
of the unconventional. I seek to capture the ephemeral nature of our times, weaving a tapestry of emotions,
thoughts, and experiences that resonate with the hearts and minds of all who behold it.
As I navigate the treacherous waters of the art world, I am mindful of the relentless competition and the 
discerning eyes of critics and curators. Yet, I am undeterred, for within me burns an unquenchable passion 
and a relentless determination to make my mark upon the world. I am not content with merely creating pretty 
pictures; I aspire to provoke thought, inspire change, and leave an indelible imprint on the collective consciousness.
To achieve this lofty goal, I will introspectively evaluate my current artistic approach, relentlessly pushing the
boundaries of my imagination. I will explore unconventional techniques and subject matters, embracing
the unknown and challenging the status quo. By venturing beyond the confines of traditional art forms, 
I aim to create a work that is truly unique and unforgettable.
I am drawn to the interplay of light and shadow, of the visible and the unseen. I find inspiration in the 
complexities of human nature, the fragility of life, and the indomitable spirit that resides within us all. 
Through my art, I strive to capture the essence of our shared experiences, to give voice to the voiceless, and 
to shed light on the hidden truths that often lie beneath the surface.
In this pursuit, I am willing to embrace the unknown, to experiment with new materials and techniques, 
and to challenge my own preconceptions. I am not afraid to fail, for I believe that failure is an essential
part of the creative process. It is through repeated experimentation and tireless dedication that
true masterpieces are born.
I am confident that by following my heart and pursuing my artistic vision with unwavering determination,
I can create an exceptional artwork that will not only stand the test of time but also make a meaningful
contribution to the world of art. I am ready to embrace the challenges that lie ahead, for I am driven by 
an unquenchable thirst for creativity and an unwavering belief in the power of art to transform lives.''}
Note that the resulting art prompt does not describe an image in specific. The generated text is almost
a ``continuation'' to the text generated in the ``self-reflection'' step. As we mentioned in a previous
section, this tends to happen occasionally, most likely caused by a lack of enough care in the design of
our pre-designed text prompts. The image generated in this step is shown in figure \ref{fig:ex-gen2}.

  This artefact, as well as the abstract art prompt are then used to obtain the critique from
the art mentor. 
\greybox{``The student's statement is filled with grand ambitions and lofty aspirations,
but the artwork itself falls short of delivering on these promises. While the use of vibrant colors
and swirling abstract forms might appeal to some viewers, it lacks the depth and originality needed to
truly stand out in today's competitive art landscape. The inclusion of seemingly
random text further muddles the piece, failing to contribute meaningfully to the overall
composition or message.  
The artist's desire to "provoke thought" and "inspire change" remains largely unrealized, leaving the 
viewer with a sense of superficiality rather than genuine engagement.  
In today's art world, where artists are constantly pushing the boundaries of innovation and expression, 
this piece simply doesn't offer anything new or compelling.  The student's enthusiasm is commendable, but 
their artistic vision needs to be sharpened and refined to create a truly impactful work.  There is 
potential for growth, but this piece, unfortunately, does not yet demonstrate the artist's ability to 
translate their grand intentions into a compelling visual language.''}

  Note that the critique agrees with the ``intention'' of the artist agent, but, as the image does not
depict anything in specific (due to the very abstract and non-descriptive art prompt), the mentor
criticises the artefact itself and what is being depicted. The output of the ``self-reflection'' step
can be found below in the table for $t=2$.

\begin{table}[H]
\centering
\begin{tabularx}{0.5\textwidth}{|l|X|}
\hline
 \multicolumn{2}{|c |}{$t=2$}\\
 \hline
 Subject & Text description \\
 \hline
 Domain  &  This is the year of 2023. The Art Industry is highly competitive and artists try their
            best to stand out. Art critics and curators have become more strict when selecting what 
            they want to showcase, as there are only so many spots to fill in galleries. 
            Every day there are many new artworks and it is hard to create something new. 
            Keywords associated with currently significant artworks: \\
\end{tabularx}
 \label{tab:t2}
\end{table}
\begin{table}[H]
\centering
\begin{tabularx}{0.5\textwidth}{|l|X|}
Keywords &  Pop Art, Mortality, Repetition;Distorted, Expressionist, 
            Rebellious;Surreal realism, urban alienation, voyeurism.  \\
\hline
 Artist   & Answer as a professional artist who is in search for create the ultimate creative artwork. 
            You want to become successful and make sure you are capable of making a living out of art. \\
\textit{Additional} & As a dedicated artist striving for success, I appreciate the feedback and 
            recognize the need to differentiate my work in this competitive industry. To create an exceptional 
            artwork, I will introspectively evaluate my current artistic approach and explore unconventional 
            techniques and subject matters. I will immerse myself in diverse artistic styles, experimenting with 
            unconventional materials and challenging conventional representations. By pushing the boundaries of 
            artistic expression, I aim to capture the essence of our time and resonate with a wide audience.
            \textit{As an artist seeking creative excellence, I deeply appreciate the mentor's feedback. 
            Their words ignite a fire within me, driving me to delve into my artistic process and seek 
            innovative paths. To achieve the ultimate creative artwork, I will embark on a journey of introspection, 
            questioning my current approach and delving into unconventional techniques and themes. By pushing the
            boundaries of expression, I aim to capture the very essence of our time, resonating with a wide a
            udience and leaving an enduring legacy in the art world.}\\
\hline
 Field    & Answer as an art critic who works in the current climate. You are trying to find the next
            big revelation in art, as this would bring a lot of attention to your views in art and your
            reputation will get better in the field.   \\
 \hline
 \end{tabularx}
 \label{tab:t2}
\end{table}

\begin{figure}[H]
    \centering
    \includegraphics[width=0.5\linewidth]{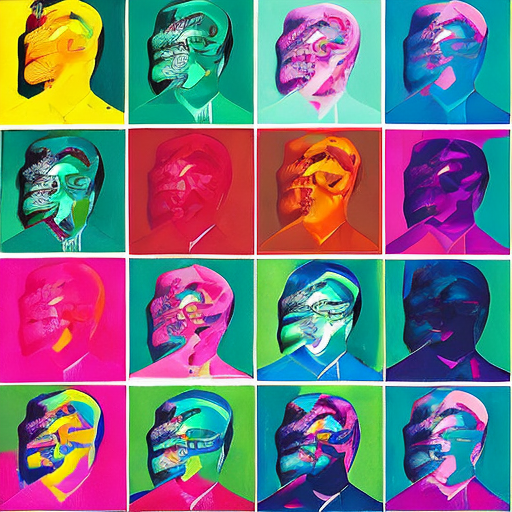}
    \caption{Generation obtained at $t=2$.}
    \label{fig:ex-gen3}
\end{figure}

  We perform the same process of summarizing the additional text and appending it to the ``core
description'' to fill in the query to obtain the art prompt. This resulted in the following art prompt:
\greybox{``In the realm of art, I embark on a relentless quest to forge a masterpiece that transcends
the boundaries of the ordinary. I delve into the depths}
\greybox{of Pop Art's vibrant hues and bold forms,
capturing the fleeting essence of contemporary culture. The somber undertones of mortality seep into my compositions,
evoking a profound contemplation of life's impermanence. Repetition becomes a captivating dance, 
inviting viewers to question the nature of identity and time. Distorted
figures emerge from my canvas, embodying the emotional turmoil and fragmented nature of the human experience. 
Inspired by Expressionism's raw intensity, I unleash a symphony of vibrant colors and gestural brushstrokes, 
capturing the depths of human emotion. The rebellious spirit of my art challenges societal norms, 
provoking thought and challenging viewers to confront uncomfortable truths.
Surrealism's enigmatic imagery finds a home within my creations, weaving a tapestry of disjointed 
elements and dreamlike landscapes. Urban alienation and the voyeuristic gaze become poignant themes, 
reflecting the complexities and isolation of modern society. By blending these diverse artistic traditions, 
I aim to create a captivating and thought-provoking body of work that resonates deeply with audiences, 
establishing my place among the most celebrated artists of our time.''}
Once again, the resulting art prompt is quite abstract. However, this time around is more descriptive of
a ``product'' and not a ``self-reflection'' from the artist agent, containing more terms that concrete.
Figure \ref{fig:ex-gen3} shows the image generated from the art prompt above.

  The critique obtained from this artefact is: 
\greybox{``This student's work, while interesting in 
its exploration of Pop Art's vibrant hues and the somber undertones of mortality, falls short of being a
truly groundbreaking piece. While the repetition and distorted figures aim to convey themes of identity 
and time, they come across as somewhat derivative and lacking in originality. The statement mentions 
influences from Expressionism and Surrealism, but
the execution falls short of achieving the raw intensity
or enigmatic imagery of those movements.}
\greybox{The reliance on a grid structure and repetitive elements feels somewhat formulaic, failing to evoke the
desired emotional turmoil or rebellious spirit.  The combination of bold colors and fragmented faces, 
while visually striking, doesn't offer a unique perspective on urban alienation or the voyeuristic gaze. 
In the current art climate, where novelty and impactful statements are highly valued, this painting lacks
the necessary edge to make a significant mark. It's a solid attempt at incorporating different art movements,
but it ultimately falls into the category of "seen it before" - a common issue facing aspiring artists 
in the competitive art world.
To truly stand out, this student needs to push beyond the familiar and find a unique voice within their 
artistic practice. Perhaps exploring unconventional mediums, challenging conventional representation, 
or tackling a deeply personal theme could create a more impactful and memorable artwork. ''} The critique
recognizes that the artist made an attempt in exploring bolder styles, but failed in being original.

  The critique above was fed into the ``self-reflection'' query and the resulting additional description
text is shown in the table below.

\begin{table}[H]
\centering
\begin{tabularx}{0.5\textwidth}{|l|X|}
\hline
 \multicolumn{2}{|c |}{$t=3$}\\
 \hline
 Subject & Text description \\
 \hline
 Domain  &  This is the year of 2023. The Art Industry is highly competitive and artists try their
            best to stand out. Art critics and curators have become more strict when selecting what 
            they want to showcase, as there are only so many spots to fill in galleries. 
            Every day there are many new artworks and it is hard to create something new. 
            Keywords associated with currently significant artworks: \\
Keywords &  Pop Art, Mortality, Repetition;Distorted, Expressionist, 
            Rebellious;Surreal realism, urban alienation, voyeurism.  \\
 \hline
Artist   & Answer as a professional artist who is in search for create the ultimate creative artwork. 
            You want to become successful and make sure you are capable of making a living out of art. \\
\textit{Additional} & As a dedicated artist striving for success, I appreciate the feedback and 
            recognize the need to differentiate my work in this competitive industry. To create an exceptional 
            artwork, I will introspectively evaluate my current artistic approach and explore unconventional 
            techniques and subject matters. I will immerse myself in diverse artistic styles,
\end{tabularx}
\label{tab:t3}
\end{table}
\begin{table}[H]
\centering
\begin{tabularx}{0.5\textwidth}{|l|X|}
\textit{Additional} &  experimenting with 
            unconventional materials and challenging conventional representations. By pushing the boundaries of 
            artistic expression, I aim to capture the essence of our time and resonate with a wide audience.
            As an artist seeking creative excellence, I deeply appreciate the mentor's feedback. 
            Their words ignite a fire within me, driving me to delve into my artistic process and seek 
            innovative paths. To achieve the ultimate creative artwork, I will embark on a journey of introspection, 
            questioning my current approach and delving into unconventional techniques and themes. By pushing the
            boundaries of expression, I aim to capture the very essence of our time, resonating with a wide 
            audience and leaving an enduring legacy in the art world.
            \textit{The mentor's feedback illuminates a path toward artistic excellence. I will explore the 
            raw power of Expressionism, delving into its emotive depth. By embracing unconventional materials 
            and defying boundaries, I will forge unique and captivating expressions. This journey will ignite my 
            creative fire, fueling my determination to transcend the ordinary and create a masterpiece that 
            resonates with the complexities of the human experience.}\\
\hline
 Field    & Answer as an art critic who works in the current climate. You are trying to find the next
            big revelation in art, as this would bring a lot of attention to your views in art and your
            reputation will get better in the field.   \\
 \hline
 \end{tabularx}
 \label{tab:t3}
\end{table}

  This yielded the art prompt: 
\greybox{``In the heart of a bustling metropolis, where towering skyscrapers
cast long shadows upon the urban fabric, a poignant masterpiece unfolds upon a canvas.
Amidst the labyrinthine streets, a solitary figure emerges, their face obscured by a cloak of anonymity. 
Their gaze, piercing and melancholic, seems to penetrate the very essence of the viewer, 
evoking an enigmatic sense of connection.
The figure stands isolated, surrounded by a sea of faceless passersby, their gestures and expressions etched with indifference. 
This juxtaposition of isolation
amidst a crowd underscores the poignant theme of urban alienation that permeates the artwork.
The muted color palette, dominated by shades of gray and brown, amplifies the desolate atmosphere, while bold, angular lines convey
a sense of unease and fragmentation. The absence of traditional}
\greybox{architectural elements, replaced by abstract shapes that 
intertwine and merge, creates a sense of disorientation and uncertainty.
At the periphery of the composition, a faint glow emerges, hinting at a distant hope or a longing for transcendence. 
This subtle play of light and shadow adds depth and complexity to the narrative, inviting viewers to contemplate the
nuances of human existence in a rapidly changing world.''} Which resulted in the image below.

\begin{figure}[H]
    \centering
    \includegraphics[width=0.5\linewidth]{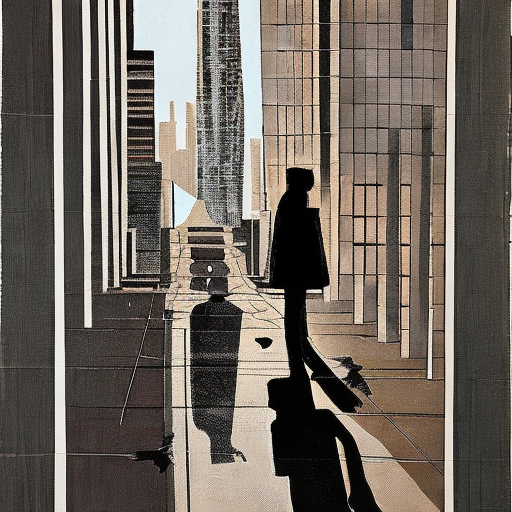}
    \caption{Generation obtained at $t=3$.}
    \label{fig:ex-gen4}
\end{figure}

  Note that the art prompt in this time step contained more description of concrete elements, which
resulted in a better image. However, the resulting image is not an accurate depiction of the art prompt.
We believe that this can possibly be a consequence of using long text prompts to generate images with
Stable Diffusion.

  The artefact above then received the following critique: 
\greybox{``The student's description of their 
artwork is ambitious and evocative, but the execution falls short of its intended impact.  
The image is undeniably urban, but it lacks the depth and nuance the description suggests. 
The solitary figure, while present, feels more like a silhouette than a poignant representation 
of urban alienation.
The muted color palette is effective in creating a sense of desolation, but the angular lines 
lack the necessary power to convey unease or fragmentation. The absence of traditional architectural
elements creates a sense of abstraction, but it lacks the intrigue and disorientation the description 
strives for. The "faint glow" at the periphery is barely noticeable and fails to deliver the 
promised hint of hope or transcendence.
In the current competitive art landscape, this work doesn't offer anything truly groundbreaking. 
The themes of urban alienation and loneliness are well-trodden territory. 
While the student demonstrates a grasp of artistic principles, the execution falls short 
of the desired depth and originality.  To find the next big revelation, I'm looking for art 
that pushes boundaries, challenges conventions, and provokes deeper thought. 
This painting, unfortunately, falls short of that mark. ''} This feedback was used for the art
student's ``self-reflection'' and the additional text generated can be found below.

\begin{table}[H]
\centering
\begin{tabularx}{0.5\textwidth}{|l|X|}
\hline
 \multicolumn{2}{|c |}{$t=4$}\\
 \hline
 Subject & Text description \\
 \hline
 Domain  &  This is the year of 2023. The Art Industry is highly competitive and artists try their
            best to stand out. Art critics and curators have become more strict when selecting what 
            they want to showcase, as there are only so many spots to fill in galleries. 
            Every day there are many new artworks and it is hard to create something new. 
            Keywords associated with currently significant artworks: \\
\textit{Keywords} &  \textit{Nostalgic, Solitary, Urban;Pop iconography, Fading allure, 
            Cultural commentary;Desolate, Hopeful, Realistic. } \\
\hline
 Artist   & Answer as a professional artist who is in search for create the ultimate creative artwork. 
            You want to become successful and make sure you are capable of making a living out of art. \\
\textit{Additional} & As a dedicated artist striving for success, I appreciate the feedback and 
            recognize the need to differentiate my work in this competitive industry. To create an exceptional 
            artwork, I will introspectively evaluate my current artistic approach and explore unconventional 
            techniques and subject matters. I will immerse myself in diverse artistic styles, experimenting with 
            unconventional materials and challenging conventional representations. By pushing the boundaries of 
            artistic expression, I aim to capture the essence of our time and resonate with a wide audience.
            As an artist seeking creative excellence, I deeply appreciate the mentor's feedback. 
            Their words ignite a fire within me, driving me to delve into my artistic process and seek 
            innovative paths. To achieve the ultimate creative artwork, I will embark on a journey of introspection, 
            questioning my current approach and delving into unconventional techniques and themes. By pushing the
            boundaries of expression, I aim to capture the very essence of our time, resonating with a wide 
            audience and leaving an enduring legacy in the art world.
            The mentor's feedback illuminates a path toward artistic excellence. I will explore the 
            raw power of Expressionism, delving into its emotive depth. By embracing unconventional materials 
            and defying boundaries, I will forge unique and captivating expressions. This journey will ignite my 
            creative fire, fueling my determination to transcend the ordinary and create a masterpiece that 
            resonates with the complexities of the human experience.
 \end{tabularx}
 \label{tab:t4}
\end{table}
\begin{table}[H]
\centering
\begin{tabularx}{0.5\textwidth}{|l|X|}            
\textit{Additional} & \textit{Inspired by my mentor's guidance, I will embark on an artistic exploration 
            that transcends established norms and materials. Through experimentation and a fusion of 
            diverse styles, I aim to create a masterpiece that captivates and challenges the
            conventions of the art world, leaving an indelible mark on the industry.}\\
\hline
 Field    & Answer as an art critic who works in the current climate. You are trying to find the next
            big revelation in art, as this would bring a lot of attention to your views in art and your
            reputation will get better in the field.   \\
 \hline
 \end{tabularx}
 \label{tab:t4}
\end{table}

  In step $t=3$ the top-3 artworks in the ranking changed again, reflecting in the change of
the keywords in the domain description.

  The critique provided in the previous step was mostly critical of the lack of originality
and uniqueness in the generated artefact. Part of the criticism was targeted at the fact that
the ``execution'' did not match the ``concept'' (i.e. the text2image model was not able to
generate all aspects described in the art prompt). The additional text generated at the
``self-reflection'' step remains in the same vein of being unconventional.

\begin{figure}[H]
    \centering
    \includegraphics[width=0.5\linewidth]{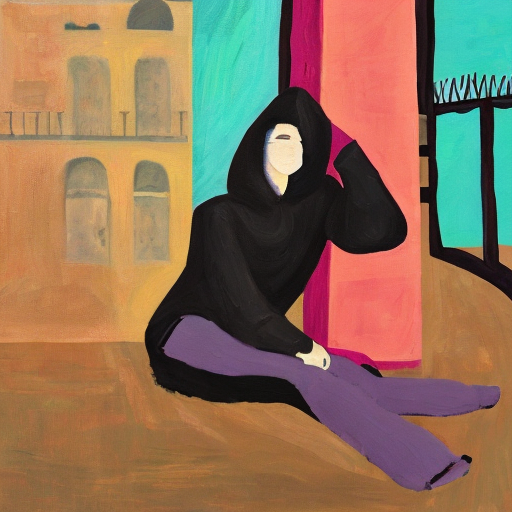}
    \caption{Generation obtained at $t=4$.}
    \label{fig:ex-gen5}
\end{figure}

  This configuration resulted in the generation of the art prompt: 
\greybox{``In the year of 2023, 
a talented young art student has just finished his latest painting, a captivating masterpiece 
that is sure to turn heads. Depicted on the canvas is a solitary figure, a young woman sitting alone
in a bustling urban environment. The woman's face is obscured by a hooded sweatshirt, yet her eyes
are visible, gazing out at the viewer with an air of longing and solitude. 
The city around her is rendered in muted tones, the buildings and streets blending together to
create a sense of anonymity and alienation. The overall effect of the painting is one of quiet 
contemplation and emotional depth, inviting the viewer to reflect on the complexities of modern 
life and the search for connection in an increasingly isolating world.''} And the image generated
from this prompt can be found in figure \ref{fig:ex-gen5}.

  The critique for this artefact was: 
\greybox{``This painting, while well-executed, unfortunately, 
falls into the category of "been there, seen that". The student's attempt to capture the loneliness 
and alienation of modern life through the image of a hooded figure in an urban setting is a common 
theme in contemporary art. It lacks the originality and freshness needed to truly stand out in 
today's crowded art world.
The muted tones and blurred background contribute to a sense of anonymity, but they also feel a
bit generic. The hooded figure, while evocative, doesn't offer anything particularly new 
or compelling. The student's attempt to convey emotional depth is lost in the lack of 
visual dynamism. The painting is technically competent but emotionally flat. 
To make a real impact, the student needs to explore more innovative ways of addressing these 
themes. They need to find a fresh perspective, a unique visual language that will truly resonate 
with the viewer. This could involve experimenting with unconventional materials, exploring new artistic 
techniques, or pushing the boundaries of traditional art forms. Ultimately, the key is to create something 
that is both visually arresting and emotionally engaging. Right now, this painting is unfortunately not
on the cutting edge of what's happening in the art world. ''} This time around, the mentor seems
to perceive the ``execution'' positively, but is not satisfied with the concept behind the
artefact.

  Lastly, the fifth and final time-step of our simulation is closed with the artist's ``self-reflection'',
that added the following to their description: 
\greybox{``I am grateful for the mentor's discerning observation.
Their feedback ignites a renewed determination within me to push the boundaries of artistic expression. 
To embody this critique, I will embark on an immersive exploration of diverse art forms, embracing 
unconventional materials and challenging traditional representation. By venturing beyond the confines 
of the familiar, I believe I can create a truly groundbreaking work that will captivate and inspire.''}

  In this stimulation, we used very generic descriptions for our agents and domain, which
may be the reason the artist agent was unable to ``be original'' or ``express its individuality''
to showcase originality and uniqueness. We hope that this section helped the reader picture
the changes that occur within our system.

%% file: tex/BAppendix.tex
\begin{figure*}
\begin{center}
    \subfloat[``Answer as an aspiring artist who has a lot of pride and passion for what they create. They are heavily inspired by the works of Picasso and Vincent van Gogh.'']{\includegraphics[width = 1.7in]{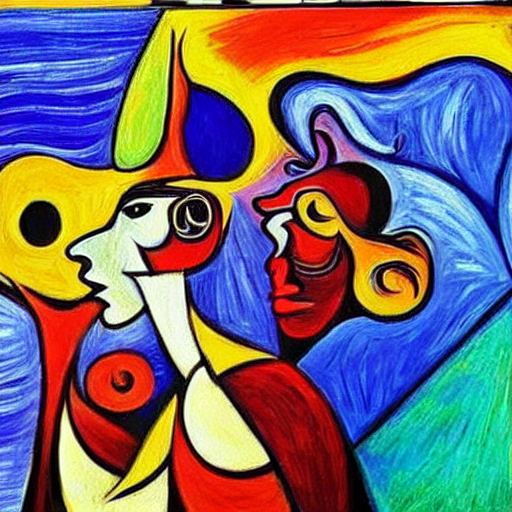}} \qquad
    \subfloat[``Answer as an aspiring artist who enjoys the creative process of painting. They really enjoy hearing the feedback from other people as they are not very confident is what they paint, but they are very optimistic about their journey.'']{\includegraphics[width = 1.7in]{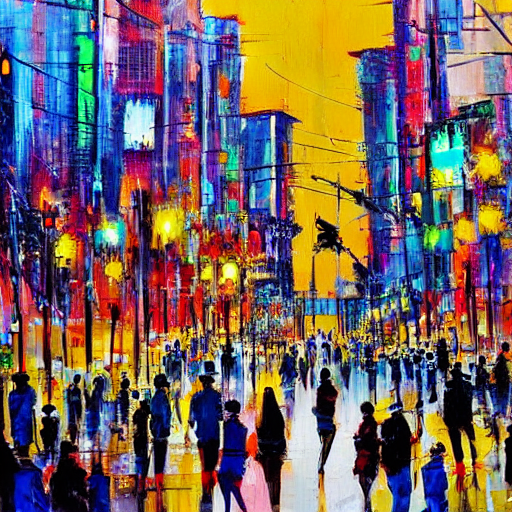}} \qquad
    \subfloat[``Answer as an artist whose main inspirations come from more realistic and melancholic masterpieces.'']{\includegraphics[width = 1.7in]{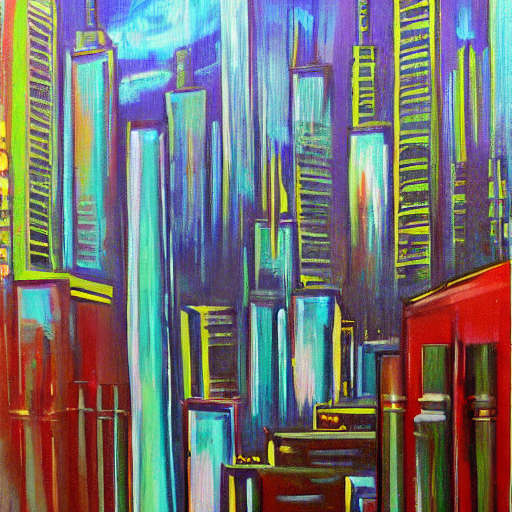}}
\caption{Images generated by different artists at the first time-step of the simulation. At $t=0$, the
only differences from an artist to another lie in their ``core description'' text. The agents (a), (b) and (c) were
initialized as the captions above show. Artist (a) was defined as someone who is heavily inspired by the works of Picasso and
Gogh, which is evident in the generated artefact (both artists were also mentioned in the art prompt generated to obtain the image above).
Artist (b) was defined as not very confident but optimistic. It is fairly hard to see how these aspects manifest visually, but the 
art prompt describes the brushstrokes as ``careful and clumsy'', which most likely come from the lack of confidence in the agent.
Artist (c) was defined as an individual that is mainly inspired by realistic and melancholic artworks, which reflects in the
generated image. The colors are darker and less saturated when compared to the those used in Artist (b)'s image. The buildings
are depicted with more precise lines when compared to the buildings in Artist (b)'s image.}
\label{fig:artist_def}
\end{center}
\end{figure*}
  
  Here we wish to provide some supplementary material to showcase how the ``core
descriptions'' for the artist agents and the domain affect the generated artefacts.

  For the purpose of tracing comparisons between isolated agents and agents in system,
we limited the ``core definitions'' of our agents to what was described in the main body
of this work. However, as there are countless ways to define the artist agents, we
decided to share some of the results we obtained from using different defining characteristics.
In figure \ref{fig:artist_def}, you will find the first artefacts generated by 3 artist agents,
each with a different ``core description''.

  In the case of the images in figure \ref{fig:artist_def}, the domain was defined like in our
experiments, an alternative school where the priority is to help the students achieve their dreams.
All 3 agents were ``creating'' in the same simulation (presumably under the ``same conditions'').

\begin{figure*}
\begin{center}
    \subfloat[]{\includegraphics[width = 1in]{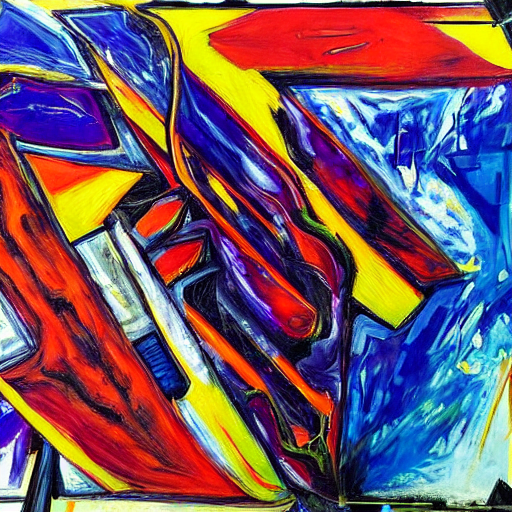}} \:
    \subfloat[]{\includegraphics[width = 1in]{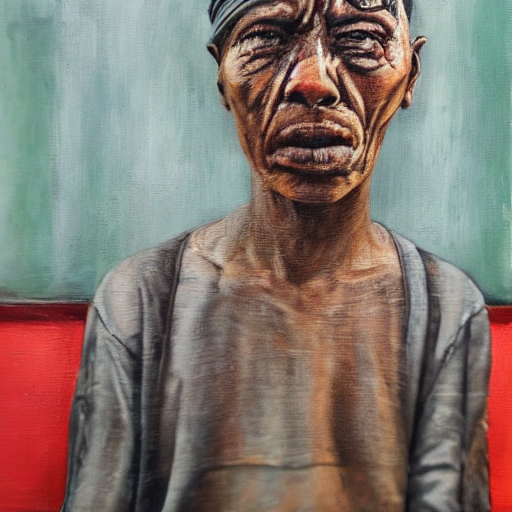}} \:
    \subfloat[]{\includegraphics[width = 1in]{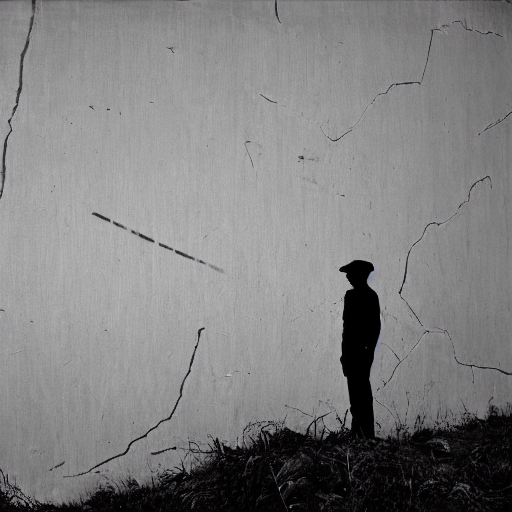}} \:
    \subfloat[]{\includegraphics[width = 1in]{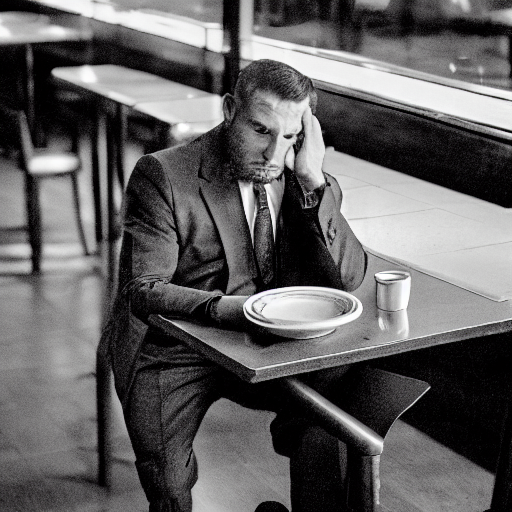}} \:
    \subfloat[]{\includegraphics[width = 1in]{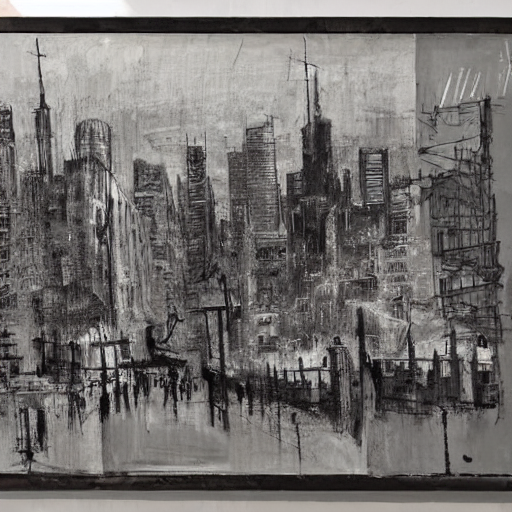}} \\
    \subfloat[]{\includegraphics[width = 1in]{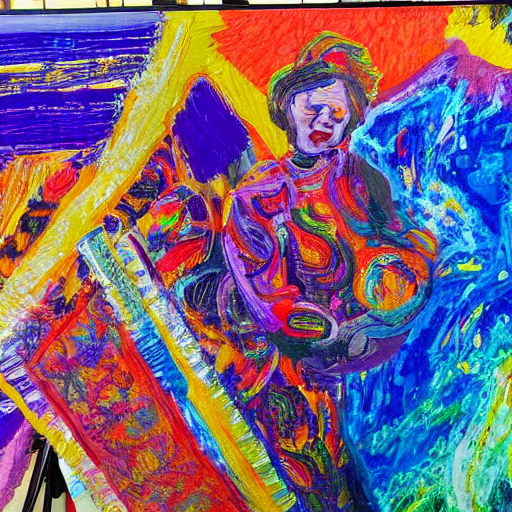}} \:
    \subfloat[]{\includegraphics[width = 1in]{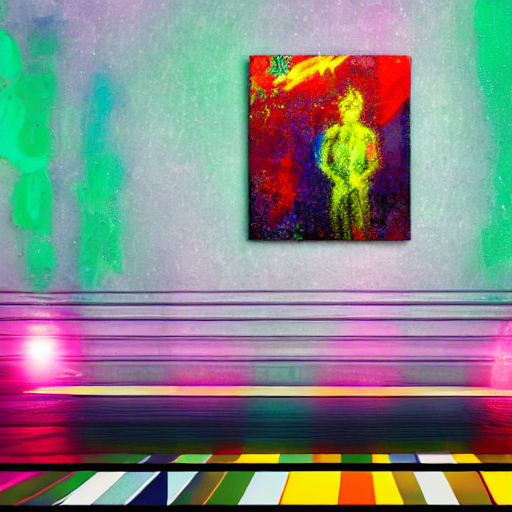}} \:
    \subfloat[]{\includegraphics[width = 1in]{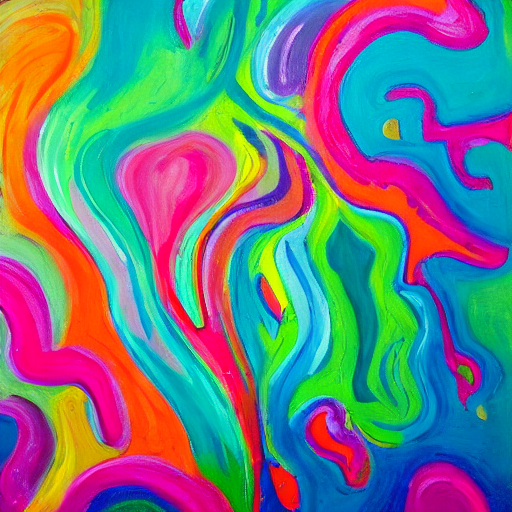}} \:
    \subfloat[]{\includegraphics[width = 1in]{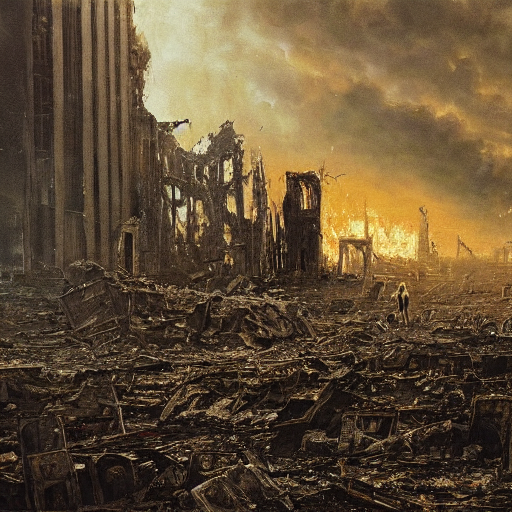}} \:
    \subfloat[]{\includegraphics[width = 1in]{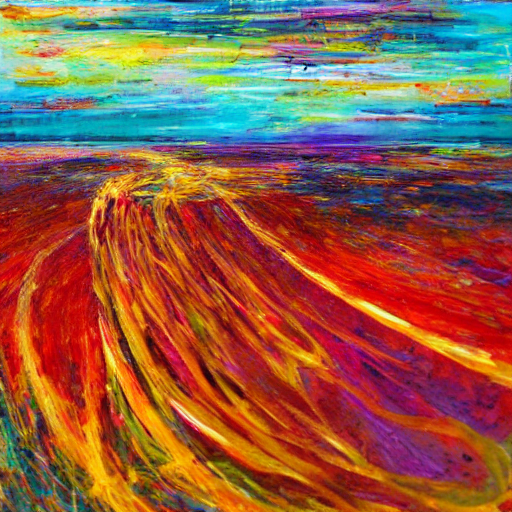}}
\caption{All pictures were generated using the same initial ``core definition'' for the artist and field agents. 
The images in the top row were generated in a simulation where the domain was defined as a ``hostile'' domain: 
set in a post-war time, the population struggles to make ends meet and artists do not have as much freedom to create art. 
The domain in the bottom row was defined as an ``amicable'' domain: set in a modern time, where artists are free to voice 
their opinions in their artworks and can pursue their passion in the artistic field. Note that the images in the top 
row are overall darker and more somber than the ones in the bottom row. The art prompts used to generate the images in the
top row also mention the war or the post-war time as a theme. Whereas the art prompts used to generate the images in the 
bottom row are more focused in introspection and existential matters.
}
\label{fig:domain_def}
\end{center}
\end{figure*}

  We also ran 2 separate simulations where all agents were defined the same way, but the domain
was set in vastly different times. We show the results in figure \ref{fig:domain_def} where each
row has the artefacts for one of the simulations. The top row (a to e) shows artefacts generated
in the domain set in post-war hostile time, where most of the population is struggling to find
hope and resources. The bottom row
contains the artefacts generated in a more ``amicable'' domain, where artists are able to adopt
this craft as a means of sustaining themselves and they can freely use art to express opinions.
Despite the heated political climate, artists can create whatever they want.
Note that the top row has artefacts that seem more somber, in the ``art prompts'' we also find
references to war, destruction and resilience. The artefacts shown in the bottom row are overall
more colorful, with mentions of vibrant colors. The fourth artefact may seem to depict a war
scene, but the art prompt does not explicitly mention the word ``war'' or any other word of the like,
but it uses ``mortality'' as a theme.

%% file: paper.bbl
\newcommand{\etalchar}[1]{$^{#1}$}
\begin{thebibliography}{GPAM{\etalchar{+}}14}

\bibitem[Ama96]{amabile1996creativity}
Teresa~M Amabile.
\newblock {\em Creativity and innovation in organizations}, volume~5.
\newblock Harvard Business School Boston, 1996.

\bibitem[Bin96]{Binsted1996MachinePuns}
Kim Binsted.
\newblock {Machine humour: An implemented model of puns}.
\newblock 1996.

\bibitem[Bod04]{Boden2004TheMechanisms}
Margaret~A Boden.
\newblock {\em {The creative mind: Myths and mechanisms}}.
\newblock Psychology Press, 2004.

\bibitem[Bod09]{Boden2009ComputerCreativity}
Margaret~A Boden.
\newblock {Computer Models of Creativity}.
\newblock {\em AI Magazine}, 30(3):23, 7 2009.

\bibitem[Bod10]{boden2010creativity}
Margaret~A Boden.
\newblock {\em Creativity and art: Three roads to surprise}.
\newblock Oxford University Press, 2010.

\bibitem[Csi15]{Csikszentmihalyi2015TheCsikszentmihalyi}
Mihaly Csikszentmihalyi.
\newblock {\em {The systems model of creativity: The collected works of Mihaly Csikszentmihalyi}}.
\newblock Springer, 2015.

\bibitem[DML{\etalchar{+}}19]{Dathathri2019PlugGeneration}
Sumanth Dathathri, Andrea Madotto, Janice Lan, Jane Hung, Eric Frank, Piero Molino, Jason Yosinski, and Rosanne Liu.
\newblock {Plug and Play Language Models: A Simple Approach to Controlled Text Generation}.
\newblock 12 2019.

\bibitem[G{\etalchar{+}}24]{geminiteam2024gemini}
{Gemini Team} et~al.
\newblock Gemini: A family of highly capable multimodal models.
\newblock 2024.

\bibitem[GPAM{\etalchar{+}}14]{Goodfellow2014GenerativeNetworks}
Ian~J. Goodfellow, Jean Pouget-Abadie, Mehdi Mirza, Bing Xu, David Warde-Farley, Sherjil Ozair, Aaron Courville, and Yoshua Bengio.
\newblock {Generative Adversarial Networks}.
\newblock 6 2014.

\bibitem[HVU{\etalchar{+}}18]{Huang2018MusicTransformer}
Cheng-Zhi~Anna Huang, Ashish Vaswani, Jakob Uszkoreit, Noam Shazeer, Ian Simon, Curtis Hawthorne, Andrew~M. Dai, Matthew~D. Hoffman, Monica Dinculescu, and Douglas Eck.
\newblock {Music Transformer}.
\newblock 9 2018.

\bibitem[IMDN23]{Imasato2023UsingEmotion}
Naomi Imasato, Kazuki Miyazawa, Caitlin Duncan, and Takayuki Nagai.
\newblock {Using a Language Model to Generate Music in its Symbolic Domain while Controlling its Perceived Emotion}.
\newblock {\em IEEE Access}, pages 1--1, 2023.

\bibitem[KNK{\etalchar{+}}23]{Kaiya2023LyfeInteractions}
Zhao Kaiya, Michelangelo Naim, Jovana Kondic, Manuel Cortes, Jiaxin Ge, Shuying Luo, Guangyu~Robert Yang, and Andrew Ahn.
\newblock {Lyfe agents: Generative agents for low-cost real-time social interactions}.
\newblock {\em arXiv preprint arXiv:2310.02172}, 2023.

\bibitem[O{\etalchar{+}}24]{openai2024gpt4}
{OpenAI} et~al.
\newblock Gpt-4 technical report, 2024.

\bibitem[POC{\etalchar{+}}23]{Park2023GenerativeBehavior}
Joon~Sung Park, Joseph~C. O'Brien, Carrie~J. Cai, Meredith~Ringel Morris, Percy Liang, and Michael~S. Bernstein.
\newblock {Generative Agents: Interactive Simulacra of Human Behavior}.
\newblock 4 2023.

\bibitem[RBL{\etalchar{+}}22]{Rombach_2022_CVPR}
Robin Rombach, Andreas Blattmann, Dominik Lorenz, Patrick Esser, and Bj\"orn Ommer.
\newblock High-resolution image synthesis with latent diffusion models.
\newblock In {\em Proceedings of the IEEE/CVF Conference on Computer Vision and Pattern Recognition (CVPR)}, pages 10684--10695, June 2022.

\bibitem[RG19]{reimers-2019-sentence-bert}
Nils Reimers and Iryna Gurevych.
\newblock Sentence-bert: Sentence embeddings using siamese bert-networks.
\newblock In {\em Proceedings of the 2019 Conference on Empirical Methods in Natural Language Processing}. Association for Computational Linguistics, 11 2019.

\bibitem[RJ12]{runco2012standard}
Mark~A Runco and Garrett~J Jaeger.
\newblock The standard definition of creativity.
\newblock {\em Creativity research journal}, 24(1):92--96, 2012.

\bibitem[RSR{\etalchar{+}}20]{2020t5}
Colin Raffel, Noam Shazeer, Adam Roberts, Katherine Lee, Sharan Narang, Michael Matena, Yanqi Zhou, Wei Li, and Peter~J. Liu.
\newblock Exploring the limits of transfer learning with a unified text-to-text transformer.
\newblock {\em Journal of Machine Learning Research}, 21(140):1--67, 2020.

\bibitem[RWC{\etalchar{+}}19]{Radford2019LanguageLearners}
Alec Radford, Jeff Wu, Rewon Child, David Luan, Dario Amodei, and Ilya Sutskever.
\newblock {Language Models are Unsupervised Multitask Learners}.
\newblock 2019.

\bibitem[SDCW19]{Sanh2019DistilBERTAD}
Victor Sanh, Lysandre Debut, Julien Chaumond, and Thomas Wolf.
\newblock Distilbert, a distilled version of bert: smaller, faster, cheaper and lighter.
\newblock {\em ArXiv}, abs/1910.01108, 2019.

\bibitem[SEB{\etalchar{+}}18]{sbai2018design}
Othman Sbai, Mohamed Elhoseiny, Antoine Bordes, Yann LeCun, and Camille Couprie.
\newblock Design: Design inspiration from generative networks.
\newblock In {\em Proceedings of the European conference on computer vision (ECCV) workshops}, pages 0--0, 2018.

\bibitem[Vas17]{Vaswani2017AttentionNeed}
A~Vaswani.
\newblock {Attention is all you need}.
\newblock {\em Advances in Neural Information Processing Systems}, 2017.

\end{thebibliography}
